\documentclass[10pt, pre, twocolumn, showpacs, eps]{revtex4-1}
\usepackage{amsmath,amssymb}
\usepackage{graphicx,color}
\usepackage{ulem}

\newcommand{\dfracp}[2]{\dfrac{\partial #1}{\partial #2}}

\newcommand{\ave}[1]{\left\langle #1 \right\rangle}
\newcommand{\oex}{\omega_{\rm ex}}

\begin{document}

\title{A role of asymmetry in linear response of globally coupled oscillator systems}
\author{Yu Terada$^{1}$}
\email{terada@sp.dis.titech.ac.jp}
\author{Keigo Ito$^{2}$}
\author{Ryosuke Yoneda$^{3}$}
\author{Toshio Aoyagi$^{4}$}
\author{Yoshiyuki Y. Yamaguchi$^{4}$}
\email{yyama@amp.i.kyoto-u.ac.jp}
\affiliation{$^{1}$Department of Mathematical and Computing Science, Tokyo Institute of Technology, 152-8552 Tokyo, Japan}
\affiliation{$^{2}$Shimadzu Corporation,
	604-8511 Kyoto, Japan}
\affiliation{$^{3}$Faculty of Engineering, Kyoto University,
        606-8501 Kyoto, Japan}
\affiliation{$^{4}$Graduate School of Informatics, Kyoto University, 
        606-8501 Kyoto, Japan}
\date{\today}
%05.45.Xt Synchronization; coupled oscillators
% 05.20.Dd Kinetic theory
% 05.70.Jk Critical point phenomena
% 74.25.N- Response to electromagnetic fields
%\today
\begin{abstract}
    The linear response is studied
    in globally coupled oscillator systems including the Kuramoto model.
    We develop a linear response theory which can be applied
    to systems whose coupling functions are generic.
    Based on the theory,
    we examine the role of asymmetry
    introduced to the natural frequency distribution,
    the coupling function, or the coupling constants.
    A remarkable difference appears in coexistence
    of the divergence of susceptibility at the critical point
    and a nonzero phase gap between the order parameter
    and the applied external force.
    The coexistence is not allowed
    by the asymmetry in the natural frequency distribution
    but can be realized by the other two types of asymmetry.
    This theoretical prediction
    and the coupling-constant dependence of the susceptibility
    are numerically verified by performing simulations
    in $N$-body systems and in reduced systems obtained
    with the aid of the Ott-Antonsen ansatz.
\end{abstract}
\maketitle

\section{Introduction}
\label{sec:introduction}

Coupled oscillator models describe
the synchronization among rhythmic elements.
A simple class of interaction is the global all-to-all couplings,
which govern dynamics through the mean field.
The Kuramoto model \cite{kuramoto-75,kuramoto-03,strogatz-00}
is a paradigmatic mean-field model,
which consists of phase oscillators having natural frequencies
and interacting with each other through
a fundamental-harmonic sine coupling function.
This model provides the synchronization transition
between the nonsynchronized state
and partially synchronized states.
The transition is continuous for the unimodal symmetric natural frequency
distributions \cite{kuramoto-75,kuramoto-84}
and can be discontinuous for bimodal symmetric ones
\cite{martens-barreto-strogatz-ott-so-antonsen-09,pazo-montbrio-09}.

The studies mentioned above
are based on the assumption of symmetry.
There is no asymmetry neither in the natural frequency
distribution nor in the odd symmetric coupling function.
However, the symmetry might not be always guaranteed in nature,
and the roles of asymmetry has to be studied accordingly.
For instance, asymmetry can modify types of transitions,
and nonstandard bifurcation diagrams were found
with asymmetric natural frequency distributions
\cite{basnarkov-urumov-07,basnarkov-urumov-08,terada-ito-aoyagi-yamaguchi-17},
and with the phase-lag parameter,
which breaks the odd symmetry of coupling function
\cite{omelchenko-wolfrum-12,omelchenko-wolfrum-13}.
Another type of asymmetry is brought by weighted-coupling constants
depending on the oscillators.
This heterogeneity induces the asymmetry in the interaction,
as a recipient and a sender are not equivalent.
Dynamics of such systems have been studied recently \cite{zhou-chen-bi-hu-liu-guan2015,xu-gao-xiang-jia-guan-zheng-16,qiu-zhang-liu-bi-boccaletti-liu-guan-15,xiao-jia-xu-lu-zheng-17}.

Asymmetry has been also investigated in the linear response to external forces.
In the Kuramoto model, the linear response was firstly derived
by using the explicit forms of stationary states
without assuming the symmetry of
the natural frequency distribution \cite{sakaguchi-88}.
According to the reported linear response formula,
one can find two remarkable phenomena,
which are the divergence of susceptibility,
and the phase gap between the order parameter and the external force.
We stress that, in the Kuramoto model,
these two phenomena never coexist.
It is impossible to observe the divergent susceptibility
with keeping the nonzero phase gap
even if the natural frequency distribution is asymmetric.
The suppression of the susceptibility is also reported in
a system with weighted-coupling constants,
where the susceptibility is constant
in the nonsynchronized state
irrespective of strength of the couplings \cite{daido-15}.

In this paper we focus on the linear response,
and study the role of asymmetry by comparing 
three types of asymmetry introduced in the Kuramoto model:
the natural frequency distribution, the coupling function,
and the coupling constants. 
Looking back to the previous works on the linear response,
some natural questions should arise:
Can we explain the above results in a unified manner?
Is it possible to have the divergence of the susceptibility
in systems with weighted-coupling constants?
Can the divergence and the phase gap coexist by introducing
asymmetry apart from the natural frequency distribution?
We will answer these questions by developing the linear response theory.

For simplifying discussions,
we concentrate on systems having only a fundamental-harmonic
sine coupling function in the main text.
In this type of systems, the linear response formula can be obtained
through the self-consistent equation for the order parameter
by using the explicit expression of stationary states
\cite{sakaguchi-88,daido-15}.
However, 
inspired by the linear response theory
in globally coupled Hamiltonian systems
\cite{patelli-gupta-nardini-ruffo-12,ogawa-yamaguchi-12},
we introduce another strategy of solving dynamics directly.
This strategy has an advantage that it can be straightforwardly extended
to systems having general coupling functions,
while the self-consistent strategy can not,
since there are several stationary states
for a given coupling function %parameter set
\cite{komarov-pikovsky-13,komarov-pikovsky-14,li-ma-li-yang-14}.
Another advantage of our strategy is that the direct analysis of the dynamics
naturally combines the linear response analysis with the stability analysis,
which is necessary to guarantee stability of reference states.

This article is organized as follows.
In Sec. \ref{sec:model} we introduce a coupled oscillator model
including the three types of asymmetry.
The linear response theory for the nonsynchronized state
is developed in Sec. \ref{sec:linear-response-formula}.
Conditions for realizing the divergence of susceptibility
and the phase gap are discussed in
Sec. \ref{sec:susceptibility}
with an explanation of the constant susceptibility
in a class of systems having weighted-coupling constants.
The linear response with each type of asymmetry
is reported in Sec. \ref{eq:response-asymmetry}
with focusing on the coexistence of the divergence and the phase gap.
Theoretical predictions are examined numerically in Sec. \ref{sec:numerics}.
The final section \ref{sec:summary} is devoted to the summary and discussions.

\section{Model}
\label{sec:model}

The phase reduction technique \cite{kuramoto-03,hoppensteadt-izhikevich-12,nakao-16} reduces a wide class of coupled limit-cycle oscillators with external forces,
and their phase dynamics are expressed by the equation
\begin{align}
\frac{d\theta_j}{dt} = \omega_j+\sum_{k=1}^{N}\Gamma_{jk}\left(\theta_j-\theta_k\right)+H_j\left(\theta_j,t\right),
\end{align}
where $\theta_j$ and $\omega_j$ are the phase 
and natural frequency of the $j$th oscillator.
We assume that $\omega_{j}$ follows
a natural frequency distribution $g(\omega)$.
The functions $\Gamma_{jk}(\theta)$ and $H_{j}(\theta,t)$,
which are $2\pi$-periodic with respect to $\theta$,
represent the interaction between the $j$th and $k$th oscillators,
and the external force applied to the $j$th oscillator,  respectively.
We note that the argument of the coupling function $\Gamma_{jk}$ is the phase difference, which is derived by the averaging method \cite{kuramoto-03,hoppensteadt-izhikevich-12,nakao-16}.

In neuronal context a neuron has specific properties for its sensitivity and interaction.
Different cells are known to exhibit various types of responses to external inputs \cite{dayan-abbott-01}.
On the other hand, as a sender of a signal, the firing of the excitatory neuron increases the potentials of other neurons while that of inhibitory decreases them.
This property is associated with positive and negative couplings with no-phase-lag sine function \cite{hong-strogatz-11}.
The heterogeneity in coupling types is ubiquitous in nature and society and it is desirable to incorporate it to a mathematical model.

\begin{figure}[h]
  \centering
	\includegraphics[scale=0.6]{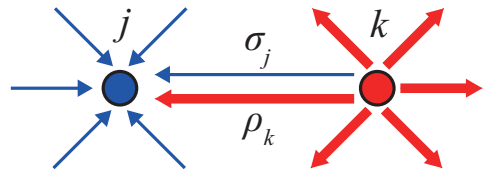}
  \caption{(Color online)
    Schematic picture of the coupling
      between the $j$th recipient and $k$th sender oscillators.
      The strength of coupling is $K\sigma_{j}\rho_{k}/N$.
  }
  \label{fig:input_output}
\end{figure}

Thus, in the main text,
we keep the above heterogeneity but restrict ourselves
to the system
\begin{equation}
  \label{eq:weighted-coupling-model}
  \frac{d\theta_{j}}{dt}
  = \omega_{j} - \frac{K}{N} \sum_{k=1}^{N} \sigma_{j} \rho_{k}
  \sin(\theta_{j}-\theta_{k}+\alpha)-h\sin(\theta_{j}-\omega_{\text{ex}}t),
\end{equation}
where the second and the third terms in the right-hand-side
represent the interaction and the external force, respectively.
The real parameter $\alpha$ $(|\alpha|<\pi/2)$
is the phase-lag parameter
\cite{sakaguchi-kuramoto-86,omelchenko-wolfrum-12,omelchenko-wolfrum-13}.
The real non-negative $h$ expresses the strength of the external force,
and the real $\oex$ is its frequency.
% The last term of the right-hand side represents the external force
% with the real strength $h$ and real frequency $\oex$.
The parameters $K$, $\sigma_{j}$ and $\rho_{k}$ are also real,
and $\sigma_{j}$ and $\rho_{k}$ determine contribution
to the coupling strength from the recipient $j$ and the sender $k$,
respectively, as shown in Fig. \ref{fig:input_output}.
These parameters give the oscillators intrinsic coupling properties and bring the heterogeneity to the network.
We can reproduce the output oriented model
$\sigma_{j}\rho_{k}=\rho_{k}$ \cite{paissan-zanette-08},
the input oriented model $\sigma_{j}\rho_{k}=\sigma_{j}$ \cite{hong-strogatz-11,iatsenko-petkoski-mcclintock-stefanovska-13},
and symmetric input-output model $\sigma_{j}\rho_{k}=\sigma_{j}\sigma_{k}$
\cite{daido-87,daido-15}.

Throughout this paper
we call the model \eqref{eq:weighted-coupling-model}
the weighted-coupling model.
The weighted-coupling model includes
the Sakaguchi-Kuramoto model by setting
$\sigma\equiv\rho\equiv 1$,
where $\sigma\equiv 1$ means $\sigma_{j}=1~(j=1,\cdots,N)$, for instance,
and the Kuramoto model by $\alpha=0$ in addition.

To measure the extent of synchrony we employ the order parameter defined by
\begin{equation}
  z  = \frac{1}{N} \sum_{j=1}^{N} e^{i\theta_{j}}.
\end{equation}
Moreover, by introducing the other order parameter 
\begin{equation}
  w = \dfrac{1}{N} \sum_{j=1}^{N} \rho_{j} e^{i\theta_{j}} ,
\end{equation}
the equation of motion \eqref{eq:weighted-coupling-model} is rewritten as
\begin{equation}
  \label{eq:weighted-coupling-model-orderparameters}
  \begin{split}
    \dfrac{d\theta_{j}}{dt}
    = \omega_{j}
    & + \frac{1}{2i} (Ke^{-i\alpha}\sigma_{j} w + he^{i\omega_{\text{ex}}t}) e^{-i\theta_{j}} \\
    & -\frac{1}{2i} (Ke^{i\alpha}\sigma_{j} \bar{w} + he^{-i\omega_{\text{ex}}t}) e^{i\theta_{j}},
  \end{split}
\end{equation}
where $\bar{w}$ is the complex conjugate of $w$.

The expression \eqref{eq:weighted-coupling-model-orderparameters} is helpful
for introducing the limit of large population, $N\to\infty$.
The conservation of the number of oscillators induces the equation of 
continuity \cite{lancellotti-05}
\begin{equation}
  \label{eq:eq_continuity}
  \dfracp{f}{t} + \dfracp{}{\theta} (vf) = 0,
\end{equation}
where $f(\theta,\omega,\sigma,\rho,t)$ is the probability density function
with the normalization condition
\begin{equation}
  \int_{0}^{2\pi} d\theta \int_{-\infty}^{\infty} d\omega\int_{-\infty}^{\infty} d\sigma \int_{-\infty}^{\infty} d\rho\,
  f(\theta,\omega,\sigma,\rho,t) = 1.
\end{equation}
The natural frequency distribution $g(\omega)$ is recovered
by integrating over $\theta,\sigma$ and $\rho$ as 
\begin{equation}
  g(\omega) = \int_{0}^{2\pi} d\theta \int_{-\infty}^{\infty} d\sigma \int_{-\infty}^{\infty} d\rho\, f(\theta,\omega,\sigma,\rho,t).\label{eq:freq_dist_integral}
\end{equation}
We note that the left-hand side of \eqref{eq:freq_dist_integral}
does not depend on the time $t$
since the natural frequency is supposed to be constant in time.
The two order parameters $z$ and $w$ are
defined by replacing the average over particles
with the average over $f$ as
%replaced by the average over $f$ as
\begin{equation}
  z(t) =  \int_{0}^{2\pi} d\theta \int_{-\infty}^{\infty} d\omega\int_{-\infty}^{\infty} d\sigma \int_{-\infty}^{\infty} d\rho\,
  f(\theta,\omega,\sigma,\rho,t) e^{i\theta}\label{eq:z_infinite}
\end{equation}
and
\begin{equation}
  w(t) = \int_{0}^{2\pi} d\theta \int_{-\infty}^{\infty} d\omega\int_{-\infty}^{\infty} d\sigma \int_{-\infty}^{\infty} d\rho\,
  f(\theta,\omega,\sigma,\rho,t) \rho e^{i\theta}.
\end{equation}
In terms of the order parameter $w$,
the velocity field $v(\theta,\omega,\sigma,t)$
of the weighted-coupling model \eqref{eq:weighted-coupling-model}
is obtained as
\begin{equation}
  \label{eq:v}
  \begin{split}
    v
    = \omega & + \dfrac{1}{2i} \left( Ke^{-i\alpha} \sigma w + he^{i\omega_{\text{ex}}t} \right) e^{-i\theta} \\
    & - \dfrac{1}{2i} \left( Ke^{i\alpha} \sigma\bar{w} + he^{-i\omega_{\text{ex}}t} \right) e^{i\theta}.
  \end{split}
\end{equation}

If the state $f(\theta,\omega,\sigma,\rho,t)$ does not depend on $\theta$,
the state is called the nonsynchronized state
and is denoted by $f_{0}(\omega,\sigma,\rho)$ throughout this paper.
The nonsynchronized state gives $w=0$,
and hence $v=\omega$ in the absence of the external force $h=0$.
It is, therefore, easy to check that the nonsynchronized state
$f_{0}(\omega,\sigma,\rho)$
is a stationary solution to the equation of continuity \eqref{eq:eq_continuity}.
In the next section \ref{sec:linear-response-formula} we linearize the equation
of continuity \eqref{eq:eq_continuity} around the nonsynchronized state $f_{0}$,
and solve it up to the leading order of a small external force $h$
to obtain the linear response.

\section{Linear response formula}
\label{sec:linear-response-formula}

\subsection{Solution to linearized equation}
\label{sec:solution}

We consider the stable nonsynchronized state $f_{0}(\omega,\sigma,\rho)$
for $t<0$ with the zero external force $h=0$,
and a small external force is turned on at $t=0$.
Due to the external force the state for $t>0$ is modified from $f_{0}$ to
\begin{equation}
  f(\theta,\omega,\sigma,\rho,t)
  = f_{0}(\omega,\sigma,\rho) + f_{1}(\theta,\omega,\sigma,\rho,t).
\end{equation}
Associated with the above expansion of $f$,
the velocity field $v$ is also expanded as
\begin{equation}
  v(\theta,\omega,\sigma,t)
  = \omega + v_{1}(\theta,\sigma,t),
\end{equation}
where
\begin{equation}
  \begin{split}
    v_{1}(\theta,\sigma,t)
    & = \dfrac{1}{2i} \left( Ke^{-i\alpha} \sigma w + h e^{i\oex t} \right) e^{-i\theta} \\
    & - \dfrac{1}{2i} \left( Ke^{i\alpha} \sigma \bar{w} + h e^{-i\oex t} \right) e^{i\theta}
  \end{split}
\end{equation}
and
\begin{equation}
  w = 
  \int_{0}^{2\pi} d\theta \int_{-\infty}^{\infty} d\omega\int_{-\infty}^{\infty} d\sigma \int_{-\infty}^{\infty} d\rho\,
  f_{1}(\theta,\omega,\sigma,\rho,t) \rho e^{i\theta}.
\end{equation}
We note that $f_{1}, v_{1}$ and $w$ come from the applied small external force,
and we may assume that they are also small.
The linearized equation is, therefore, obtained as
\begin{equation}
  \label{eq:Extended-linear-eq}
  \dfracp{f_{1}}{t} + \omega \dfracp{f_{1}}{\theta}
  + f_{0}(\omega,\sigma,\rho) \dfracp{v_{1}}{\theta} = 0.
\end{equation}
As $f_{1}$ is small, the order parameter $z$,
\begin{equation}
  \label{eq:z-def}
  z = 
  \int_{0}^{2\pi} d\theta \int_{-\infty}^{\infty} d\omega\int_{-\infty}^{\infty} d\sigma \int_{-\infty}^{\infty} d\rho\,
  f_{1}(\theta,\omega,\sigma,\rho,t) e^{i\theta},
\end{equation}
is also small. 
Our job is to calculate $z(t)$ for large $t$.
% the asymptotic \blue{dynamics} of $z$ for a given small external force.

To solve the linearized equation \eqref{eq:Extended-linear-eq},
we perform the Fourier series expansion with respect to $\theta$
and the Laplace transform with respect to $t$.
From the expression of $z$ \eqref{eq:z-def},
we can find that $z(t)$ is recovered from the Fourier $-1$ mode of $f_{1}$.
Correspondingly, we focus on the external force of the Fourier $-1$ mode,
which is denoted by
\begin{equation}
  H(t)=he^{i\oex t}\Theta(t) 
\end{equation}
with the unit step function $\Theta(t)$.
After some calculations described
in Appendix \ref{sec:generalized-linear-response},
%and \ref{sec:analytic-continuation},
for $f_{1}(\theta,\omega,\sigma,\rho,0)=0$,
the Laplace transform of $z(t)$, denoted by $\hat{z}(s)$,
is formally given by
\begin{equation}
  \label{eq:hatzs}
  \hat{z}(s) = \chi(s) \hat{H}(s), 
\end{equation}
where
\begin{equation}
  \label{eq:chis}
   \chi(s) = F(s) + Ke^{-i\alpha} \dfrac{F_{\sigma}(s) F_{\rho}(s)}{D_{K,\alpha}(s)},
\end{equation}
and
the functions $F_{X}(s)$ and $D_{K,\alpha}(s)$ are defined by
\begin{equation}
  \label{eq:FXs}
  F_{X}(s) = \pi \int_{L} d\omega\int_{-\infty}^{\infty} d\sigma \int_{-\infty}^{\infty} d\rho\,
  \dfrac{Xf_{0}(\omega,\sigma,\rho)}{s-i\omega},
\end{equation}
and
\begin{equation}
  \label{eq:D}
  D_{K,\alpha}(s) = 1 - Ke^{-i\alpha} F_{\sigma\rho}(s).
\end{equation}
The subscript $X$ is $X\in\{1,\sigma,\rho,\sigma\rho\}$
and we used the simple notation of $F(s)=F_{1}(s)$.
The functions are defined in the region ${\rm Re}\,s>0$
to ensure convergence of the Laplace transform
(see \eqref{eq:Laplace-transform})
and the integral contour $L$ with respect to $\omega$
runs on the real axis.
However, the functions are analytically continued to the whole complex $s$ plane
by smoothly modifying the contour $L$
to avoid the singularity at $\omega=-is$
as shown in Appendix \ref{sec:analytic-continuation}.

The formula \eqref{eq:hatzs} is the base of the following discussions.
This formula can be extended to general coupling functions
and to general external forces beyond
 the fundamental-harmonics function
as shown in Appendix \ref{sec:generalized-linear-response}.

\subsection{Linear response and susceptibility}
\label{sec:linear-response}

Temporal evolution of the order parameter $z(t)$ is obtained by
performing the inverse Laplace transform of $\hat{z}(s)$ as
\begin{equation}
  \label{eq:inverse-Laplace-transform}
  z(t) = \dfrac{1}{2\pi i} \int_{\Gamma}ds\, \hat{z}(s) e^{st}.
\end{equation}
The inverse Laplace transform picks up the singularities of $\hat{z}(s)$.
More precisely, if $\hat{z}(s)$ has a simple pole at $s=s_{0}$, 
then $z(t)$ has the mode of $\exp(s_{0}t)$.
Keeping this fact in mind we consider asymptotic behavior of $z(t)$.

We assumed that $f_{0}$ is stable,
and hence, no singularity of $\hat{z}(s)$ appears
in the domain ${\rm Re}\,s>0$.
The poles in the domain ${\rm Re}\,s<0$
give exponentially decreasing modes.
Therefore, if there are singularity points
on the imaginary axis ${\rm Re}\,s=0$,
the asymptotic behavior of $z(t)$ is dominated by them. 
Let us consider possible sources of imaginary singularities
by recalling \eqref{eq:hatzs} and \eqref{eq:chis}.
We can say that the functions $F_{X}(s)$ has basically
no singularity on the imaginary axis as the result of the analytic continuation.
%\textcolor{red}{for the following reason}.
%although they have poles on real axis for a Lorentzian \red{$g(\omega)$}.
The roots of $D_{K,\alpha}(s)$ are possibly on the imaginary axis,
but they accidentally appear for special values of $K$,
as we have to determine the two parameters $K$ and the pure imaginary $s$
to satisfy the two conditions ${\rm Re}\,D_{K,\alpha}(s)=0$
and ${\rm Im}\,D_{K,\alpha}(s)=0$.
%is accidental as at the critical point $K_{\rm c}$.
Consequently, the remaining source of singularities
on the imaginary axis is the Laplace transform of the external force,
$\hat{H}(s)$.

The Laplace transform of the external force $H(t)=he^{i\oex t}\Theta(t)$ is
written as
\begin{equation}
  \hat{H}(s) = \dfrac{h}{s-i\oex},
\end{equation}
and hence, the asymptotic behavior of $z(t)$ is expressed as
\begin{equation}
  z(t) \xrightarrow{t\to +\infty} e^{i\oex t} \chi(i\oex) h
\end{equation}
in the linear regime.
% , where
% \begin{equation}
%   \label{eq:chis}
%    \chi(s) = F(s) + Ke^{-i\alpha} D_{K,\alpha}^{-1}(s) F_{\sigma}(s) F_{\rho}(s).
% \end{equation}
Moving to the rotating frame, the constant asymptotic response is obtained as
\begin{equation}
  \label{eq:chi_rot}
  e^{-i\oex t} z(t) \xrightarrow{t\to +\infty} \chi(i\oex) h.
\end{equation}
From the above discussions, we call $\chi(s)$ the susceptibility here.
We remark that the susceptibility is invariant under the exchange of
the input parameter $\sigma$ and the output one $\rho$
from the formula \eqref{eq:chis}.

\section{Analysis of susceptibility}
\label{sec:susceptibility}

\subsection{Phase gap and divergence of susceptibility}
\label{sec:divergence-phasegap}

The susceptibility formula \eqref{eq:chis} provides two notable phenomena:
the phase gap and the divergence of the susceptibility.

The phase gap refers to the disagreement of the phases
of the external force and the responded order parameter
in the rotating frame with the frequency $\oex$.
In \eqref{eq:chi_rot}, $h$ is positive real,
hence the nonzero phase gap occurs if and only if
\begin{equation}
  \label{eq:phase-gap-condition}
    {\rm Im}\,\chi(i\oex)\neq 0
  \quad \text{or} \quad
  \chi(i\oex)<0. 
\end{equation}
% \blue{hence the phase gap is zero if and only if
% \begin{equation}
%   \label{eq:zero-phase-gap-condition}
%   \chi(i\oex) \in \mathbb{R}_{+},
% \end{equation}
% where $\mathbb{R}_{+}$ is the set of positive real numbers.
% }

% The other is the divergence of susceptibility $\chi(i\oex)$.
% From the susceptibility formula \eqref{eq:chis}, % with \eqref{eq:chi_rot},
% the divergence %of susceptibility $\chi(i\oex)$
% occurs when
The divergence of the susceptibility $\chi(i\oex)$ occurs,
from the susceptibility formula \eqref{eq:chis}, when
\begin{equation}
  \label{eq:divergence-condition-0}
  D_{K,\alpha}(i\oex)=0 
\end{equation}
with the collateral condition
\begin{equation}
  \label{eq:collateralconditions}
  F_{\sigma}(i\oex)F_{\rho}(i\oex)\neq 0,
\end{equation}
since the continued functions $F_{X}(s)$'s have no divergence.
The real and imaginary parts of the condition
\eqref{eq:divergence-condition-0}
give $K$ and $\oex$, respectively.
Indeed, $\oex$ is determined by the imaginary part
\begin{equation}
  \label{eq:ImD-criticalpoint}
  {\rm Im} \left[ e^{-i\alpha} F_{\sigma\rho}(i\oex)\right] = 0,
\end{equation}
which does not depend on $K$, and then $K$ is given, with this $\oex$, from the real part
\begin{equation}
  \label{eq:Kc}
  1 - K {\rm Re}\left[ e^{-i\alpha} F_{\sigma\rho}(i\oex)\right] = 0.
\end{equation}
If ${\rm Re}[e^{-i\alpha}F_{\sigma\rho}(i\oex)]\neq 0$,
we can take the real parameter $K$ satisfing \eqref{eq:Kc},
and therefore, the divergence condition is reduced
to selecting $\oex$ which satisfies \eqref{eq:ImD-criticalpoint}.

In the Kuramoto model ($\alpha=0, \sigma\equiv\rho\equiv 0$)
with symmetric $g(\omega)$,
the zero external frequency $\oex=0$ satisfies
the imaginary part \eqref{eq:ImD-criticalpoint}
and the real part \eqref{eq:Kc} gives
\begin{equation}
  K = \dfrac{2}{\pi g(0)}.
\end{equation}
This value agrees with the synchronization transition point
as long as $g(\omega)$ is symmetric and unimodal \cite{kuramoto-03}.
The distributions $g(\omega)$ used in Sec. \ref{sec:numerics}
are also unimodal and the pair $(\oex,K)$ satisfying
the condition \eqref{eq:divergence-condition-0} is unique.
Thus, we call $K$ determined by the condition \eqref{eq:Kc}
as the critical point and denote it by $K_{\rm c}$
in the following discussions.

\subsection{Constant susceptibility in nonsynchronized state}
\label{sec:constant-susceptiblity}

Before progressing to the comparison of the three types of asymmetry,
we explain and generalize the constant susceptibility reported in \cite{daido-15}.
As in \cite{daido-15},
we assume that $\sigma$ and $\rho$ are independent from $\omega$.
The nonsynchronized state is then written as
\begin{equation}
  \label{eq:independence}
  f_{0}(\omega,\sigma,\rho)
  = \dfrac{g(\omega)}{2\pi} P(\sigma,\rho).
\end{equation}
This decomposition simplifies the function $F_{X}(s)$ as
\begin{equation}
  \label{eq:FXs-simplified}
  F_{X}(s) = \ave{X}_{\sigma,\rho} F(s),
\end{equation}
where
\begin{equation}
  \ave{X}_{\sigma,\rho} = \int_{-\infty}^{\infty} d\sigma \int_{-\infty}^{\infty} d\rho
  \,X P(\sigma,\rho).
\end{equation}
Therefore, the susceptibility \eqref{eq:chis} is also simplified as
\begin{equation}
  \label{eq:chis-independent}
  \chi(s) = \left[ 1
    + \dfrac{\ave{\sigma}_{\sigma,\rho}\ave{\rho}_{\sigma,\rho}}{\ave{\sigma\rho}_{\sigma,\rho}}
    \dfrac{1-D_{K,\alpha}(s)}{D_{K,\alpha}(s)} \right] F(s).
  % \blue{
  % \chi(s) = \left[
  %   1 - KF(s) \left( \ave{\sigma\rho}_{\sigma,\rho} - \ave{\sigma}_{\sigma,\rho} \ave{\rho}_{\sigma,\rho} \right) \right] \dfrac{F(s)}{D_{K,\alpha}(s)}.
  % }
\end{equation}

Let us assume $\ave{\sigma}_{\sigma,\rho}=0$ or $\ave{\rho}_{\sigma,\rho}=0$.
In this case the formula \eqref{eq:chis-independent} immediately
gives the constant susceptibility
\begin{equation}
  \chi(i\oex) = F(i\oex)
\end{equation}
in the nonsynchronized state, where we see that the right-hand side does not depend on the coupling strength.
In \cite{daido-15}, $\ave{\sigma}_{\sigma,\rho}=0$ is assumed,
and the constant susceptibility is a consequence of this assumption.
We note that $\rho\equiv\sigma$ is also assumed in \cite{daido-15},
and the finite critical point $K_{\rm c}$ exists from \eqref{eq:Kc},
since $\ave{\sigma\rho}_{\sigma,\rho}=\ave{\sigma^{2}}_{\sigma,\rho}$ is positive
unless $\sigma\equiv 0$.

We give two remarks for the independent case \eqref{eq:independence}.
First, the constant susceptibility is a special case,
since $\chi$ may diverge if
$\ave{\sigma}_{\sigma,\rho}\ave{\rho}_{\sigma,\rho}\neq 0$.
Second, the imaginary part of the divergence condition
\eqref{eq:ImD-criticalpoint}
implies that $F(i\oex)$ is real.
Thus, $\chi(i\oex)$ is also real and %\textcolor{red}{the}
one of the nonzero phase gap condition,
${\rm Im}\chi(i\oex)\neq 0$, is not satisfied.

% Second, the coexistence of the phase gap and the divergence is impossible
% for $\alpha=0$ \blue{unless $\chi$ is real negative,}
% \textcolor{magenta}{\it (I am not sure if $\chi$ can be negative for $|K|<|K_{\rm c}|$, while we can say $F(i\oex),D_{K,\alpha}(i\oex)>0$.)}
% since \blue{the imaginary part \eqref{eq:ImD-criticalpoint}
% induces real $F(i\oex)$ for any $K$,
% and $D_{K,0}(i\oex)$ and $\chi(i\oex)$ are}
% also real accordingly.

% Another remark is on the coexistence of the phase gap
% and the divergence.
% The divergence condition \eqref{eq:divergence-condition} and $\alpha=0$
% induces a real $F(i\oex)$ since $D_{K,0}(s)=1-K\ave{\sigma\rho}_{\sigma,\rho}F(s)$.
% Therefore, $\chi(i\oex)$
% is real and the phase gap condition \eqref{eq:phase-gap-condition}
% is not satisfied.
% Consequently, the coexistence is impossible in the independent case.

\section{Linear response with asymmetry}
\label{eq:response-asymmetry}

We investigate the role of asymmetry in the linear response
through the phase gap and the divergence of susceptibility.
Asymmetry is introduced
into the natural frequency distribution $g(\omega)$,
the coupling function along with the phase-lag parameter $\alpha$,
or the coupling constants $K\sigma_{j}\rho_{k}$.
Each type of asymmetry is studied without external forces in the Kuramoto model
($\sigma\equiv\rho\equiv 1, \alpha=0$) \cite{kuramoto-03,basnarkov-urumov-08},
in the Sakaguchi-Kuramoto model
($\sigma\equiv\rho\equiv 1, \alpha\neq 0$) \cite{sakaguchi-kuramoto-86,omelchenko-wolfrum-12},
and in the frequency-weighted-coupling model
($\rho\equiv 1, \omega\equiv\sigma, \alpha=0$) \cite{xu-gao-xiang-jia-guan-zheng-16,qiu-zhang-liu-bi-boccaletti-liu-guan-15},
respectively.
In the last model, 
%\textcolor{red}{when studying the linear response the case}
the case $\rho\equiv 1$ is equivalent to the case $\sigma\equiv 1$
in the linear response
%\textcolor{red}{to the case} $\sigma\equiv 1$
due to the exchange symmetry
between $\sigma$ and $\rho$ in the susceptibility $\chi(s)$ \eqref{eq:chis}.
The relation $\omega\equiv\sigma$ is introduced
to break the independence \eqref{eq:independence},
which gives rise to ${\rm Im}\,\chi(i\oex)=0$,
and the presented form of the correlation is not essential.

% In the subsection \ref{sec:divergence-phasegap}
% we arrange the conditions to have the phase gap and the divergence.
The susceptibilities in the three models
are given in the subsection \ref{sec:susceptibility-in-three-models}.
The coexistence of the two phonemena is discussed
in the subsection \ref{sec:coexistence}.

\subsection{Susceptibility in the three models}
\label{sec:susceptibility-in-three-models}

The three models have the constant parameter $\rho\equiv 1$.
Due to this constant parameter, the susceptibility is simplified as
\begin{equation}
  \label{eq:simple-chis}
  \chi(s) = \dfrac{F(s)}{D_{K,\alpha}(s)}.
%  = \dfrac{F(s)\overline{D_{K,\alpha}(s)}}{|D_{K,\alpha}(s)|^{2}}.
\end{equation}
The function $D_{K,\alpha}(s)$ is written as
\begin{equation}
  D_{K,\alpha} = 1 - K e^{-i\alpha} F_{Y}(s),
\end{equation}
where $Y=1$ in the Kuramoto model and the Sakaguchi-Kuramoto model,
and $Y=\sigma$ in the frequency-weighted-coupling model.
% \begin{equation}
%   D_{K,\alpha}(s) = \left\{
%     \begin{array}{ll}
%       1 - Ke^{i\alpha} F(s) & (\text{in K, SK models}) \\
%       1 - Ke^{i\alpha} F_{\sigma}(s) & (\text{in RC model}) \\
%     \end{array}
%   \right.
% \end{equation}
% By using the relation $Ke^{-i\alpha}F_{Y}(s)=1-D_{K,\alpha}(s)$,
% the susceptibility $\chi(s)$ is simplified to
% \blue{
% The simplified susceptibility \eqref{eq:simple-chis} modifies
% the phase gap condition \eqref{eq:phase-gap-condition} as
% \begin{equation}
%   \label{eq:phase-gap-condition-simple}
%   {\rm Im}\left[ \overline{D_{K,\alpha}(i\oex)} F(i\oex) \right]
%   \neq 0,
% \end{equation}
% and the divergence condition \eqref{eq:divergence-simple} as
% \begin{equation}
%   {\rm Im}\left[ e^{-i\alpha} F_{Y}(i\oex) \right] = 0.
% \end{equation}
% }

At the pure imaginary point $s=i\oex$,
the functions $F(s)$ and $F_{\sigma}(s)$ take the values
\begin{equation}
  \label{eq:Fiex}
  F(i\oex) = \dfrac{\pi}{2} g(\oex) + \dfrac{i}{2} {\rm PV}
  \int_{-\infty}^{\infty}d\omega \dfrac{g(\omega)}{\omega-\oex},
\end{equation}
and
\begin{equation}
  \label{eq:Fsigmaiex}
  F_{\sigma}(i\oex) = \dfrac{\pi}{2} \oex g(\oex) + \dfrac{i}{2} {\rm PV}
  \int_{-\infty}^{\infty}d\omega \dfrac{\omega g(\omega)}{\omega-\oex}.
\end{equation}
where PV represents the Cauchy principal value.
The expression of $F_{\sigma}(i\oex)$ is obtained from the correlation form
$\sigma=\omega$ in the frequency-weighted-coupling model.

% \textcolor{red}{In particular, for the frequency-weighted-coupling model $F_{\sigma}(s)$ is expressedv similarly as
% \begin{equation}
%   \label{eq:Fsigmaiex}
%   F_{\sigma}(i\oex) = \dfrac{\pi}{2} \oex g(\oex) + \dfrac{i}{2} {\rm PV}
%   \int_{-\infty}^{\infty}d\omega \dfrac{\omega g(\omega)}{\omega-\oex},
% \end{equation}
% where we use the correlation form $\sigma=\omega$.}

\subsection{Coexistence of phase gap and divergence}
\label{sec:coexistence}

Let us assume the divergence of the susceptibility
\eqref{eq:ImD-criticalpoint}, 
that is, $D_{K,\alpha}(i\oex)$ is real. 
This condition
% \textcolor{red}{, that is,} the condition
% \eqref{eq:divergence-condition} \textcolor{red}{holds}, which
is equivalent to
\begin{equation}
  \label{eq:divergence}
  {\rm Im}\left[ e^{-i\alpha} F_{Y}(i\oex) \right] = 0,
\end{equation}
%\textcolor{red}{which means} $D_{K,\alpha}(i\oex)$ is real.
and simplifies the former sufficient condition of \eqref{eq:phase-gap-condition} for the nonzero phase gap into
\begin{equation}
  \label{eq:phasegap}
  {\rm Im}\,F(i\oex) \neq 0.
\end{equation}
We examine whether the phase gap condition \eqref{eq:phasegap}
can hold under the divergence condition \eqref{eq:divergence}.
In the followings,
we assume that the support of $g(\omega)$ is the whole real axis.
%\textcolor{red}{below}.
% The divergence and the phase gap can coexist
% if the two conditions \eqref{eq:divergence} and \eqref{eq:phasegap}
% are not mutually exclusive.

In the Kuramoto model, we set $\alpha=0$ and $Y=1$.
The divergence condition \eqref{eq:divergence} becomes
\begin{equation}
  \label{eq:divergence-K}
  {\rm Im}\,F(i\oex) = 0,
\end{equation}
and also from the condition \eqref{eq:Kc} the critical point $K_{\rm c}$ is given by
\begin{equation}
  \label{eq:Kc-K}
  K_{\rm c} = \dfrac{1}{{\rm Re}F(i\oex)} = \dfrac{2}{\pi g(\oex)}.
\end{equation}
Obviously, the divergence condition \eqref{eq:divergence-K}
and the phase gap condition \eqref{eq:phasegap} are mutually exclusive.
The other possibility for nonzero phase gap
is that $\chi(i\oex)$ is negative real. However,
$F(i\oex)=\pi g(\oex)/2$ is positive real
and $D_{K,0}(i\oex)=1-KF(i\oex)$ is also positive real for $K<K_{\rm c}$.
Therefore, the susceptibility $\chi(i\oex)$ is positive real
and the two phenomena never coexist.
This result is consistent with the previous work
by the self-consistent analysis \cite{sakaguchi-88}.

In the Sakaguchi-Kuramoto model, we set $Y=1$ again but $\alpha\neq 0$.
The divergence condition \eqref{eq:divergence} is read as
\begin{equation}
  \label{eq:divergence-SK}
  {\rm Im}\,F(i\oex) \cos\alpha - {\rm Re}\,F(i\oex) \sin\alpha = 0,
\end{equation}
which gives the critical point as
\begin{equation}
  \label{eq:Kc-SK}
    K_{\rm c} = \dfrac{2\cos\alpha}{\pi g(\oex)}.
\end{equation}
The condition \eqref{eq:divergence-SK} implies
\begin{equation}
  {\rm Im}\,F(i\oex) = {\rm Re}\,F(i\oex) \tan\alpha
  = \dfrac{\pi}{2} g(\oex) \tan\alpha,
\end{equation}
and hence, the phase gap condition \eqref{eq:phasegap} holds
for $\alpha\neq 0$.

% the phase-lag parameter $\alpha$ and the natural frequency distribution
% $g(\omega)$
% satisfy
% This condition is satisfied by $\alpha\neq 0$ %\blue{\sout{,\pi ({\rm mod} 2\pi)}}$
% and $g(\omega)$ whose support is the real axis.
%for the phase-lag parameter satisfying $\tan\alpha\neq 0$
% [i.e. $\alpha\neq0,\pi\,(\mathrm{mod }2\pi)$]
% and the natural frequency distribution $g(\omega)$
% whose support is the real axis.

Finally, in the frequency-weighted-coupling model,
we set $\alpha=0$ but $Y=\sigma$.
We have the divergence condition \eqref{eq:divergence} of the form
\begin{equation}
  \label{eq:divergence-FWC}
  0 = {\rm Im}\,F_{\sigma}(i\oex)
  = \dfrac{1}{2} + \oex {\rm Im}\,F(i\oex).
\end{equation}
This condition implies that $\oex\neq 0$ holds for the divergence, and 
is compatible with the phase gap condition \eqref{eq:phasegap} as
\begin{equation}
  {\rm Im}\,F(i\oex) = - \dfrac{1}{2\oex} \neq 0.
\end{equation}
We note that the critical point is given by
\begin{equation}
  \label{eq:Kc-FWC}
  K_{\rm c} = \dfrac{2}{\pi \oex g(\oex)}.
\end{equation}

From the above discussions,
we conclude that the disagreement of the two phases
of $F(i\oex)$ and $e^{-i\alpha}F_{Y}(i\oex)$
is essential to realize the coexistence.
The two quantities are identical in the Kuramoto model
($\alpha=0, Y=1$)
even if the natural frequency distribution $g(\omega)$ is asymmetric,
% \textcolor{red}{even with the} asymmetry of the natural frequency distribution $g(\omega)$,
and therefore, the coexistence is impossible.
However, the phase-lag parameter $\alpha$
or the difference between $F$ and $F_{Y}$ permits the coexistence.

\section{Numerical simulations}
\label{sec:numerics}

We numerically examine theoretical predictions described
%in Secs. \ref{sec:linear-response-formula}
%\ref{sec:divergence-phasegap} 
in Sec. \ref{eq:response-asymmetry}.

\subsection{Family of natural frequency distributions}

For considering both symmetric and asymmetric natural frequency distributions,
we introduce a family of $g(\omega)$ as in \cite{terada-ito-aoyagi-yamaguchi-17}:
\begin{equation}
  \label{eq:family}
  g(\omega) = \dfrac{c}{[(\omega-\Omega)^2+\gamma_{1}^{2}][(\omega+\Omega)^{2}+\gamma_{2}^2]},
  % \quad (\Omega,\gamma_{1},\gamma_{2}>0),
\end{equation}
where $\Omega\geq 0, \gamma_{1},\gamma_{2}>0$ and
the normalization constant $c$ is given by
\begin{equation}
  c = \dfrac{\gamma_{1}\gamma_{2}[(\gamma_{1}+\gamma_{2})^{2}+4\Omega^{2}]}{\pi(\gamma_{1}+\gamma_{2})}.
\end{equation}
Using the scaling of the variables,
we may set $\gamma_{2}=1$ without loss of generality.
%and hence we simply denote $\gamma_{1}$ by $\gamma$.
Moreover, we may concentrate on the region $\gamma_1\leq 1$
by considering the replacement of $\theta\to -\theta$.
The distribution is symmetric if $\gamma_1=1$ or $\Omega=0$
and tends to be bimodal with large $\Omega$.
%\textcolor{red}{The phase diagram for the Kuramoto model without external force}

\begin{figure}[h]
  \centering
	\includegraphics[scale=0.46]{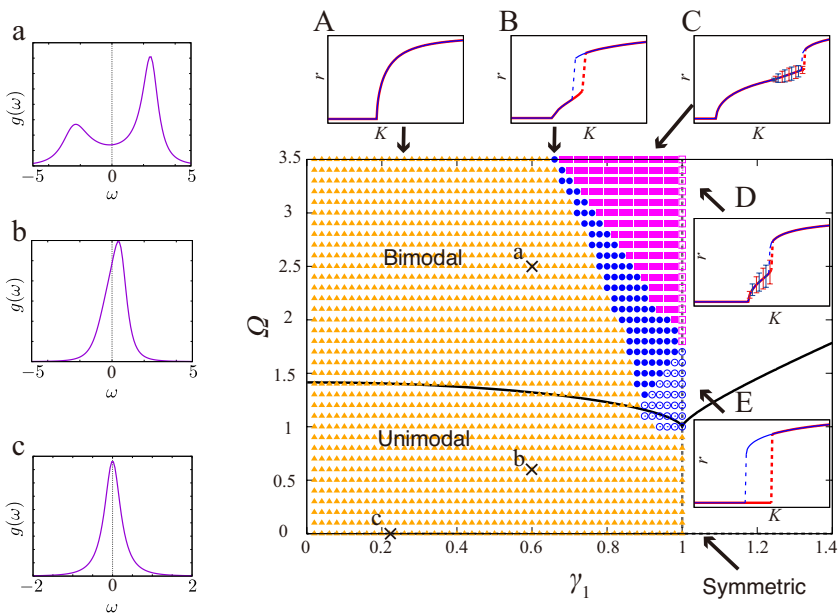}
  \caption{(Color online)
    Parameter space
    for the family \eqref{eq:family} with $\gamma_{2}=1$.
    The solid black line is the border between the unimodal and bimodal regions.
    The distribution is symmetric on the line $\gamma_{1}=1$ or $\Omega=0$.
    Three representative forms of $g(\omega)$ are shown in the left panels 
    indicated by a, b and c for the corresponding points, respectively,
    where the points b and c in the unimodal side are
    used in the numerical examinations. 
    The five domains are A (orange filled triangle),
      B (blue filled circle), C(magenta filled rectangle),
      D (magenta open rectangle) and E (blue open circle).
      Each inset indicated by A, B, C, D and E shows a schematic bifurcation diagram for the Kuramoto model
      in the indicated domain, where the vertical bars in C and D represent
      the standard deviation of $r(t)$.
      }
  \label{fig:phase-diagram}
\end{figure}

To capture the parameter dependence in the family \eqref{eq:family},
we compute the bifurcation diagram for a give set of parameters
$(\gamma_{1},\Omega)$
in the reduced system for the Kuramoto model, which is derived
by using the Ott-Antonsen ansatz \cite{ott-antonsen-08,ott-antonsen-09}
(see Appendix \ref{sec:reduction} for the derivation).
The parameter space $(\gamma_{1},\Omega)$ is roughly divided
into five domains in the computed range as shown in Fig. \ref{fig:phase-diagram}:
% The parameter space is shown in Fig. \ref{fig:phase-diagram}, where we plot 5 different points denoting roughly the kinds of the bifurcation in the Kuramoto model.
In the domain A the system undergoes only the continuous transition.
The domain B represents the continuous and successive discontinuous transitions.
The domains C and D include the oscillations before the discontinuous transition, where the continuous transition occur in C while it does not in D.
In the domain E the system has only the discontinuous transition.
The thick red lines are obtained by increasing $K$ whereas
the thin blue  lines by decreasing it.
%This is obtained by integrating the reduced system,
%which is derived
%This parameter space includes the regions with the two nonstandard bifurcation diagrams (B and C)
Two remarks are as follows. % \textcolor{red}{following}.
First, the two nonstandard bifurcation diagrams
reported in \cite{terada-ito-aoyagi-yamaguchi-17}
appear in the domains B and C,
which are unveiled by introducing the asymmetry of $g(\omega)$. 
Second, the discontinuous transition can occur
in the asymmetric unimodal distributions,
which will be discussed in the last section.

To examine the susceptibility around the critical point,
we select the continuous transition region.
Moreover, we choose the unimodal distributions for simplicity:
an asymmetric point $(\gamma_{1},\Omega)=(0.6,0.6)$ for the Kuramoto model,
and a symmetric point $(0.25,0)$ for the Sakaguchi-Kuramoto model
and the frequency-weighted-coupling model.
Stability analysis is described in Appendix \ref{sec:Nyquist} and
we confirm that the nonsynchronized state is unstable for $K>K_{\rm c}$,
where $K_{\rm c}$ is given by \eqref{eq:Kc-K}, \eqref{eq:Kc-SK},
or \eqref{eq:Kc-FWC}.

Numerical examinations are performed by $N$-body simulations
and by the reduced system.
Temporal evolution is computed by using the fourth-order
Runge-Kutta algorithm with the time step $\Delta t=$0.1.

\onecolumngrid

\twocolumngrid

\subsection{The Kuramoto model}
\label{sec:kuramoto_simulation}

If the natural frequency distribution $g(\omega)$ is symmetric and unimodal,
then the divergence condition ${\rm Im}\,F(i\oex)=0$
is satisfied if and only if $\oex=0$
as shown in Appendix \ref{sec:symmetry-case}.
In this case the divergence of the susceptibility occurs at the critical point
$K_{\rm c}=2/[\pi g(0)]$ but the phase gap is zero.

\begin{figure}[h]
  \centering
	\includegraphics[scale=0.65]{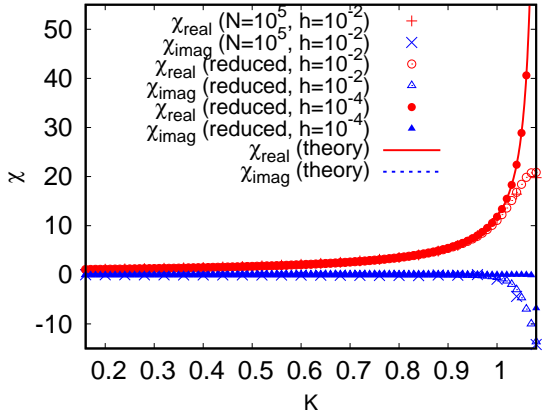}
  \caption{(Color online)
    Susceptibility in the Kuramoto model 
    with an asymmetric natural frequency distribution,
    $(\gamma_1,\Omega)=(0.6,0.6)$.
    %\textcolor{red}{We set}
    The numerical simulations are conducted with $N=10^{5}$, $h=10^{-2}$ and $10^{-4}$.
    The frequency of the external force is set
    as $\omega_{\text{ex}}=0.303819$,
    which induces the zero imaginary part of the susceptibility.
    The divergence of $\chi(i\oex)$ is observed at the critical point
    $K=K_{\rm c}\simeq 1.084618$, the right boundary of the panel,
    but no phase gap appears for sufficiently small $h$.
    %except for a neighborhood of the critical point.
    %See the text for details.
    }
  \label{fig:susceptibility_cor_sym_omega_ex}
\end{figure}

A similar thing happens for an asymmetric $g(\omega)$
with $(\gamma_{1},\Omega)=(0.6,0.6)$.
The divergence remains by seeking the value $\oex\simeq 0.303819$
which satisfies the condition \eqref{eq:divergence-K}
at the critical point $K_{\rm c}\simeq 1.084618$, \eqref{eq:Kc-K},
%\eqref{eq:ImD-criticalpoint}, ${\rm Im}F(i\oex)=0$,
while the phase gap vanishes.
This theoretical prediction is successfully confirmed in 
Fig. \ref{fig:susceptibility_cor_sym_omega_ex},
if the external force $h$ is sufficiently small,
although the zero phase gap is sensitive for the strength
of the external force near the critical point.

% On the other hand, if $g(\omega)$ is asymmetric,
% ${\rm Im} \,F(0)$ is not necessarily zero for $\oex=0$.
% Therefore, the phase gap is not zero in general,
% but the divergence of $\chi(0)$ disappears.
% This prediction is verified %for $(\gamma_{1},\Omega)=(0.6,0.6)$
% in Fig. \ref{fig:susceptibility_cor_sym}.

\begin{figure}
  \centering
  \includegraphics[scale=0.65]{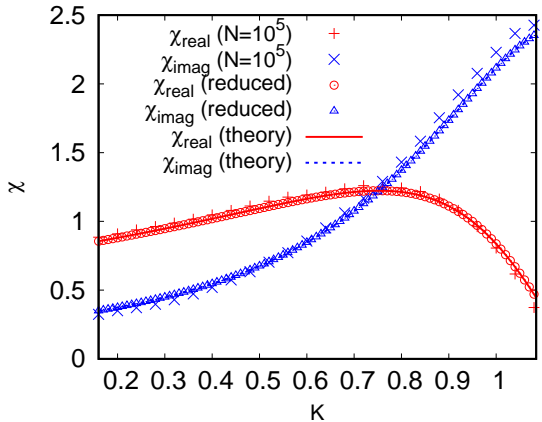}
  \caption{(Color online)
    Susceptibility in the Kuramoto model 
    with an asymmetric natural frequency distribution,
    where $(\gamma_1,\Omega)=(0.6,0.6)$.
    %\textcolor{red}{We set} 
    The numerical simulations are conducted with $N=10^{5}$ and $h=10^{-2}$. 
     The frequency of the external force is set to zero.
    The phase gap appears as the nonzero imaginary part of $\chi$
    but there is no divergence at the critical point
    $K=K_{\text{c}}\simeq 1.084618$,
    which is the right boundary of the panel.}
  \label{fig:susceptibility_cor_sym}
\end{figure}

In contrast, when we break the divergence condition
\eqref{eq:divergence-K} by choosing $\oex=0$,
the phase gap is not zero but the divergence of the susceptibility disappears.
This behavior is verified in Fig.\ref{fig:susceptibility_cor_sym}.

\subsection{The Sakaguchi-Kuramoto model}

\begin{figure}[h]
  \centering
	\includegraphics[scale=0.65]{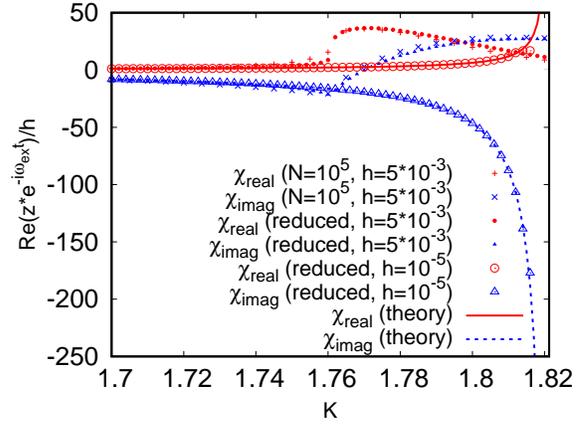}
  \caption{(Color online)
    Susceptibility in the Sakaguchi-Kuramoto model
    with $(\gamma_1,\Omega)=(0.25,0)$.
    The numerical simulations are conducted with $N=10^{5}$ and $h=5\times10^{-3},10^{-5}$. 
    %\textcolor{red}{We set }
    %$h=10^{-5}$.
    The frequency of external force is set as $\oex=1$,
    and the phase-lag parameter as $\alpha=-1.436475$
    to satisfy the divergence condition \eqref{eq:divergence-SK}.
    The critical point is $K_{\rm c}\simeq 1.821283$.}
    %\textcolor{magenta}{{\it What happens if $h=10^{-2}$ or $10^{-3}$?}} }
      % We then set $\omega_{\text{ex}}=1$, which induces the zero imaginary part
      % of the susceptibility.}
  \label{fig:susceptibility_sk_omega_ex}
\end{figure}

The discussion in Sec. \ref{sec:coexistence} says
that the nonzero phase gap and the divergence of the susceptibility coexists
under nonzero phase-lag parameter.
We use a symmetric unimodal $g(\omega)$ with $(\gamma_{1},\Omega)=(0.25,0)$.
To set $\omega_{\text{ex}}=1$ satisfying
the divergence condition \eqref{eq:divergence-SK},
we choose the phase-lag parameter as $\alpha=-1.436475$.
The critical point \eqref{eq:Kc-SK} is $K_{\rm c}\simeq 1.821283$.
% \begin{equation}
%   K_{\rm c}
%   = \dfrac{2\cos\alpha}{\pi g(\oex)}
%   = 1.821283.
% \end{equation}
Under this setting, the coexistence is observed in 
Fig. \ref{fig:susceptibility_sk_omega_ex}
for sufficiently small $h$.
%\textcolor{red}{for small value of $h$}.
We note that if $h$ is not small enough the deviation
from the theoretical values is not negligible.
In fact, the deviation is observed with $h=5\times10^{-3}$
in the $N$-body and reduced systems.
The deviated response suggests a kind of bifurcation
with respect to the strength of external force, $h$.
Studying this deviation is interesting but out of range
of this article, since our main topic is the linear response.

%We used small value of $h$,
%since the type of transition changes with rather large $h$.
%\textcolor{red}{In fact, with $h=5\times10^{-3}$ the $N$-body and reduced systems exhibit the deviations from the theoretical values, which are caused by the large value of $h$.} 

% \red{
% \subsection{The Sakaguchi-Kuramoto model}
% \label{sec:sk_simulation}}

% \red{
% Next, we investigate the Sakaguchi-Kuramoto model with the nonzero phase-lag as in Sec. \ref{sec:ex_skmodel}.
% The phase-lag parameter brings the asymmetric interaction between a pair of the oscillators, and this asymmetry permits the coexistence of the phase  gap and the divergence, although the asymmetry of the natural frequency distribtuion in the Kuramoto model does not.}

% \red{
% As a example, we choose the distribution parameter set as $(\gamma_1,\Omega)=(0.25,0)$ generating the symmetric distribution, and the phase-lag value as $\alpha=-1.436475$.
% For the susceptibility $\chi$ to diverge we set the frequency of the external force as $\omega_{\text{ex}}=0.303819$ from the condition \eqref{eq: divergence_sk}.
% In this case we can calculate the critical value from $K_{\text{c}}=2\cos\alpha/\left[\pi g\left(\omega_{\text{ex}}\right)\right]$ and $K_{\text{c}}=1.821283$.}
% \red{Then , the real and imaginary parts of the susceptibility are shown in Fig. \ref{fig:susceptibility_sk_omega_ex}, which is obtained from the the reduced simulation with $h=10^{-5}$ and our theory.
% In the Sakaguchi-Kuramoto model we confirm that the asymmetry by the phase lag causes the coexistence of the nonzero phase gap and divergence at the critical point.}

\subsection{The frequency-weighted-coupling model}

We finally investigate the frequency-weighted-coupling model,
where the coupling parameters are set as
$\sigma\equiv\omega$ and $\rho\equiv 1$.
%$\sigma_j=\omega_j$ and $\rho_k\equiv1$}.
The natural frequency distribution $g(\omega)$ is again taken
at the point $(\gamma_{1},\Omega)=(0.25,0)$,
and the phase-lag parameter is zero, $\alpha=0$.
The external frequency $\oex$ is determined from the divergence
condition \eqref{eq:divergence-FWC} as $\oex=0.5$.
The critical point \eqref{eq:Kc-FWC} is calculated as $K_{\rm c}=5$.
% \begin{equation}
%   K_{\rm c} = \dfrac{2}{\pi\oex g(\oex)} = 5.
% \end{equation}

\begin{figure}[h]
  \centering
	\includegraphics[scale=0.65]{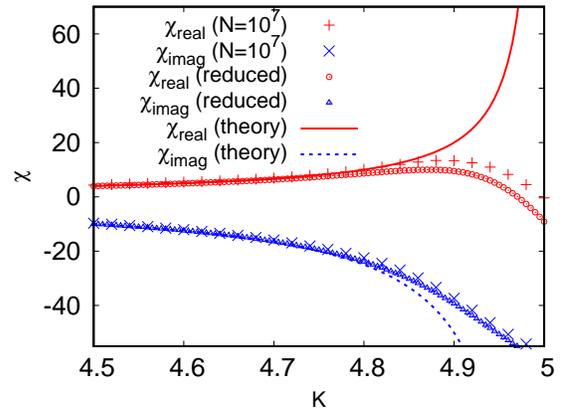}
  \caption{(Color online)
    Susceptibility in the frequency-weighted-coupling model
    with $(\gamma_{1},\Omega)=(0.25,0)$.
    The numerical simulations are conducted with $N=10^{7}$ and $h=10^{-3}$.
    The phase-lag parameter is set to zero.
    The frequency of the external force is given by $\oex=0.5$,
    which satisfies the divergence condition \eqref{eq:divergence-FWC}.
    The critical point is $K_{\rm c}=5$.
    }
    % \red{The real and imaginary parts of the susceptibility in the generalized model with $(\gamma_1,\Omega)=(0.25,0)$ and $\omega_{\text{ex}}=0.5$. This parameter set induces the coexistence of the phase gap and the divergence of susceptibility. While the nonzero phase gap can be seen easily, the divergence at the critical point is not conspicuous because of the finite value of $h$.}}
  \label{fig:susceptibility_cor_sym_real_imag}
\end{figure}

The susceptibility $\chi(i\oex)$ is exhibited in
Fig. \ref{fig:susceptibility_cor_sym_real_imag}.
The theoretical curves imply the coexistence of the divergence
of the susceptibility and the nonzero phase gap,
but the numerically obtained values are not in good agreement with
the theoretical curves near the critical point.
We have two sources of this discrepancy, which are the finite-size effects
and the finiteness of $h$,
as observed in the Kuramoto model.
We note that $h$ must be larger than the finite-size fluctuation,
which may be of $O(1/\sqrt{N})$,
to correctly pick up the linear response.
The strength $h=10^{-3}$ is close to the boundary with $N=10^{7}$
in Fig. \ref{fig:susceptibility_cor_sym_real_imag},
and hence, we can not use smaller $h$.

\begin{figure}[h]
  \centering
	\includegraphics[scale=0.65]{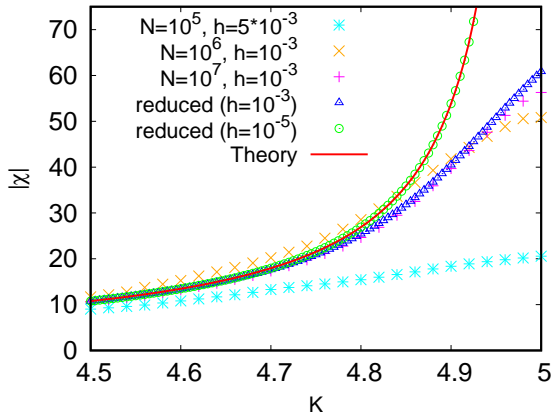}
  \caption{(Color online)
    $N$-dependence and $h$-dependence of the absolute values
    of the susceptibility in the frequency-weighted-coupling model
    with the parameter set $(\gamma_{1},\Omega)=(0.25,0)$.}
    % \red{The absolute of the susceptibility in the generalized model with $(\gamma_1,\Omega)=(0.25,0)$ and $\omega_{\text{ex}}=0.5$, which is the same as in Fig. \ref{fig:susceptibility_cor_sym_real_imag}. The large sizes of the system in $N$-body simulation improve the correspondence with the theoretical result. In the simulation of the reduced equations, the smaller value of $h$ exhibits the better illustration of the divergence.}}
  \label{fig:susceptibility_cor_sym_ansatz}
\end{figure}

Based on the above discussion, to verify the theoretical prediction,
%To investigate the two sources,
we computed the $N$-dependence and $h$-dependence
of the absolute value of the susceptibility in 
Fig. \ref{fig:susceptibility_cor_sym_ansatz}.
First, as $N$ increases with a fixed $h$,
the $N$-body simulations approaches to the reduced system,
which corresponds to the large population limit $N\to\infty$.
Thus, the reduced system must be useful with smaller $h$.
Second, the reduced system approaches to the theoretical curve as $h$ goes to $0$.
We, therefore, conclude that the divergence of the susceptibility
appears and it can coexist with the nonzero phase gap
if the coupling parameter $\sigma$
correlates with the natural frequency $\omega$.

\begin{figure}[h]
	\centering
	\includegraphics[scale=0.65]{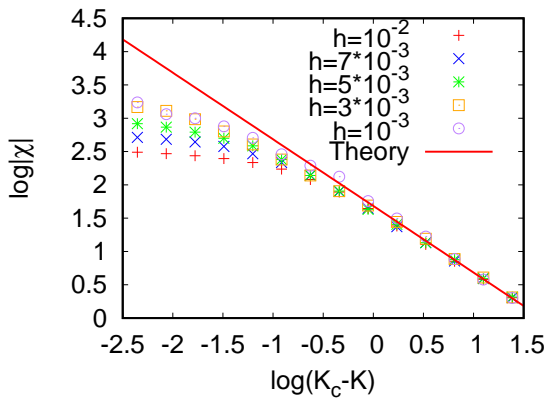}
  \caption{(Color online)
    Log-log plot of the absolute value of the susceptibility
    $|\chi(i\oex)|$ against $K_{\rm c}-K$. 
    We set the system size as $N=10^{5}$.
    The straight red line is obtained by the theory and has the slope $-1$.}
    % \red{The susceptibility with $(\gamma_1,\Omega)=(0.25,0)$, which are used in Figs. \ref{fig:susceptibility_cor_sym_real_imag}, \ref{fig:susceptibility_cor_sym_ansatz} and \ref{fig:susceptibility_cor_sym_phase_gap}. The points denote the results by the $N$-body simulation with $N=10^5$ and the line expresses the theoretical one. The slope of the theoretical line is $1$.}}
  \label{fig:susceptibility_cor_sym_log}
\end{figure}

The divergence is characterized by the critical exponent $\gamma$,
defined by
\begin{equation}
  |\chi(i\oex)|\propto |K_{\rm c}-K|^{-\gamma}
\end{equation}
near the critical point.
The critical exponent is obtained as $\gamma=1$
in Fig. \ref{fig:susceptibility_cor_sym_log},
which reports the convergence of the numerical points to the theoretical curve
in the limit $h\to 0$.
We note that the critical exponent $\gamma=1$ is also obtained
by the self-consistent analysis \cite{sakaguchi-88}
and by the finite-size scaling \cite{hong-chate-tang-park-15}
in the Kuramoto model.

% The convergence in the limit $h\to 0$ is also observed
% in Fig. \ref{fig:susceptibility_cor_sym_log},
% which indicates that the critical exponent is $\gamma=1$,
% where $\gamma$ is defined by
% \begin{equation}
%   |\chi(i\oex)|\propto |K_{\rm c}-K|^{-\gamma}
% \end{equation}
% near the critical point.

\section{Summary and discussions}
\label{sec:summary}

We studied the role of asymmetry in coupled oscillator systems
with shedding light on the linear response in the nonsynchronized state.
Three types of asymmetry are considered, which appear
in the natural frequency distribution,
in the coupling function,
or in the coupling constants.
The linear response is theoretically derived 
by directly solving the equation of continuity up to the linear order
of a small external force.
To compare the three types, we focus on the coexistence
of the phase gap and the divergence of the susceptibility.
The asymmetry in the natural frequency distribution
does not permit the coexistence,
but the other two types of asymmetry do.
Asymmetry in the natural frequency distribution
and in the coupling function provides similar nonstandard
bifurcation diagrams
\cite{omelchenko-wolfrum-13,terada-ito-aoyagi-yamaguchi-17}.
However, the two types of asymmetry are not mutually substitutable
from the view point of the linear response.
This result is helpful to identify an unknown system from the linear response,
as the system must be beyond the description of the Kuramoto model
when the coexistence is observed.

In a weighted-coupling model with the random distribution of the coupling parameters
the constant susceptibility has been reported in \cite{daido-15}.
Using the proposed linear response theory,
we revealed that the constant susceptibility is realized
under a special setting,
and that the divergence of the susceptibility is possible in general.

These theoretical predictions, and the susceptibility itself,
are verified by performing numerical simulations of $N$-body dynamics
and of the reduced systems introduced by the Ott-Antonsen ansatz.
The numerical computations suggest that we have to pay attention
to the strength of the external force,
since a small but rather large external force can bring a finite
phase gap even if the system setting theoretically requires the zero phase gap.

The linear response theory for the first-harmonic coupling function
is straightforwardly extended to general coupling functions,
when we consider the nonsynchronized state.
This point should be stressed as an advantage of our strategy.
However, the linear response theory in the partially synchronized states
has not been obtained along our line,
and it must be useful for physical applications.
Another interesting extension is to systems on networks
beyond the all-to-all connection.

Finally, in the Kuramoto model,
the asymmetry in the natural frequency distribution
produces discontinuous transitions even when a distribution is unimodal.
Our finding is that the discontinuity occurs with smooth distributions, while
non-smooth distributions are known to cause the discontinuous transitions
\cite{basnarkov-urumov-08}.
We should study how the asymmetry generates the discontinuous synchronization transition with 
smooth unimodal distributions.

\acknowledgements
Y.T. is supported by MEXT KAKENHI Grant Number 17H00764.
T.A. is supported by MEXT KAKENHI Grant Numbers 15H05877 and 26120006, 
and by JSPS KAKENHI Grant Numbers 16KT0019,15587273 and 15KT0015.
Y.Y.Y. acknowledges the support of KAKENHI Grant Number 16K05472.

\appendix

\section{Linear response formula in generalized system}
\label{sec:generalized-linear-response}

In the main text we specifies the coupling function
as only the fundamental-harmonic sine function,
as in Eq. \eqref{eq:weighted-coupling-model}.
However, our linear response theory is not restricted to
this type of coupling function,
and we here derive the expression of the susceptibility
in more general systems.

We generalize the model as
\begin{equation}
  \label{eq:generalized-model}
  \frac{d\theta_{j}}{dt}
  = \omega_{j} + \dfrac{K}{2i N} \sum_{k=1}^{N}
  \sigma_{j} \rho_{k} \Gamma(\theta_{j}-\theta_{k})
  + \frac{1}{2i} H\left(\theta_{j},t\right),
\end{equation}
where $\Gamma(\theta)$ is the coupling function
and $H(\theta,t)$ represents the external force.
The factor $1/2i$ is multiplied for the later convenience, and is not essential.
The equation of continuity is written as
\begin{equation}
  \label{eq:generalized-eq-continuity}
  \dfracp{f}{t} + \dfracp{}{\theta}(vf) = 0,
\end{equation}
where the velocity field $v\left(\theta,\omega,\sigma,t\right)$ is defined by
% The generalization of model induces the modification of the velocity
% field $v$ as
\begin{equation}
  \label{eq:generalized-velocity-field-1}
  \begin{split}
    & v\left(\theta,\omega,\sigma,t\right)
    = \omega + \dfrac{1}{2i} H\left(\theta,t\right) \\
    & + \dfrac{K\sigma}{2i} 
    \int_{0}^{2\pi}d\theta'
    \int_{-\infty}^{\infty} d\omega 
    \int_{-\infty}^{\infty} d\rho
    \,\rho \Gamma\left(\theta-\theta'\right) f\left(\theta',\omega,\sigma,\rho,t\right).
  \end{split}
\end{equation}
As in Sec.\ref{sec:linear-response-formula},
we expand $f$ around the nonsynchronized stationary state
$f_{0}(\omega,\sigma,\rho)$ as $f=f_{0}+f_{1}$,
where $f_{1}(\theta,\omega,\sigma,\rho,t)$ is regarded as a small deviation.

Let us introduce the Fourier series expansions
\begin{equation}
  \Gamma(\theta) = \sum_{n-\infty}^{\infty} \tilde{\Gamma}(n) e^{in\theta},
  \quad
  H(\theta,t) = \sum_{n-\infty}^{\infty} \tilde{H}(n,t) e^{in\theta},
\end{equation}
and
\begin{equation}
  f_{1}(\theta,\omega,\sigma,\rho,t)
  = \sum_{n-\infty}^{\infty} \tilde{f}_{1}(n,\omega,\sigma,\rho,t) e^{in\theta}.
\end{equation}
The Laplace transform of $Y(t)$ is defined by
\begin{equation}
  \label{eq:Laplace-transform}
  \hat{Y}(s) = \int_{0}^{\infty}dt\,Y(t) e^{-st},
  \quad {\rm Re}\,s > 0.
\end{equation}
The condition ${\rm Re}\,s>0$ is introduced to ensure the convergence of the integral.
Performing the Fourier-Laplace transform,
we have the Laplace transform of $\tilde{f}_{1}$ as
\begin{equation}
  \label{eq:hatf}
  \begin{split}
    & \hat{f}_{1}(-n,\omega,\sigma,\rho,s)
    = \dfrac{\tilde{f}_{1}(-n,\omega,\sigma,\rho,0)}{s-in\omega} \\
    & + \left[ K\sigma \tilde{\Gamma}(-n) \hat{w}_{n}(s) + \hat{H}(-n,s) \right]
  \dfrac{nf_{0}(\omega,\sigma,\rho)}{2(s-in\omega)},
  \end{split}
\end{equation}
where $\hat{w}_{n}(s)$ is the Laplace transform of
\begin{equation}
  \begin{split}
    w_{n}(t)
    & = \int_{0}^{2\pi} d\theta
    \int_{-\infty}^{\infty} d\omega
    \int_{-\infty}^{\infty} d\sigma
    \int_{-\infty}^{\infty} d\rho
    \, \rho e^{in\theta} f(\theta,\omega,\sigma,\rho,t) \\
    & = 2\pi \int_{-\infty}^{\infty} d\omega
    \int_{-\infty}^{\infty} d\sigma
    \int_{-\infty}^{\infty} d\rho
    \,\rho \tilde{f}_{1}(-n,\omega,\sigma,\rho,t).
  \end{split}
\end{equation}
Another family of order parameters $z_{n}(t)$ is similarly defined by
\begin{equation}
  \begin{split}
    z_{n}(t)
    & = \int_{0}^{2\pi} d\theta
    \int_{-\infty}^{\infty} d\omega
    \int_{-\infty}^{\infty} d\sigma
    \int_{-\infty}^{\infty} d\rho
    \, e^{in\theta} f(\theta,\omega,\sigma,\rho,t) \\
    & = 2\pi \int_{-\infty}^{\infty} d\omega
    \int_{-\infty}^{\infty} d\sigma
    \int_{-\infty}^{\infty} d\rho
    ~ \tilde{f}_{1}(-n,\omega,\sigma,\rho,t).
  \end{split}
\end{equation}

Multipling \eqref{eq:hatf} by $2\pi\rho$
and integrating over $\omega,\sigma$ and $\rho$,
we have the self-consistent equation for $\hat{w}_{n}$ as
\begin{align}
  \hat{w}_{n}(s) =& G_{\rho}(n,s)
  + K\tilde{\Gamma}(-n) F_{\sigma\rho}(n,s) \hat{w}_{n}(s)\notag\\
  &+ F_{\rho}(n,s) \hat{h}(-n,s).
\end{align}
The functions $F_{X}(n,s)$ and $G_{X}(n,s)$ are defined by
\begin{equation}
  \label{eq:FXns}
  F_{X}(n,s) = \pi \int_{L} d\omega
  \int_{-\infty}^{\infty} d\sigma
  \int_{-\infty}^{\infty} d\rho
  \dfrac{n X f_{0}(\omega,\sigma,\rho)}{s-in\omega},
\end{equation}
and
\begin{equation}
  \label{eq:GXns}
  G_{X}(n,s) = 2\pi \int_{L} d\omega
  \int_{-\infty}^{\infty} d\sigma
  \int_{-\infty}^{\infty} d\rho
  \dfrac{X \tilde{f}_{1}(-n,\omega,\sigma,\rho,0)}{s-in\omega}.
\end{equation}
The formal solution of the Laplace transform $\hat{w}_{n}$ is written as
\begin{equation}
  \label{eq:wns}
  \hat{w}_{n}(s) = \dfrac{1}{D_{K}(n,s)} \left[
    G_{\rho}(n,s) + F_{\rho}(n,s) \hat{h}(-n,s) \right],
\end{equation}
where
\begin{equation}
  D_{K}(n,s) = 1 - K\tilde{\Gamma}(-n) F_{\sigma\rho}(n,s).
\end{equation}

As done for $\hat{w}_{n}(s)$,
the Laplace transform $\hat{z}(s)$ is solved by multiplying
\eqref{eq:hatf} by $2\pi$
and integrating over $\omega,\sigma$ and $\rho$.
The solution is found as
\begin{equation}
  \hat{z}_{n}(s)
  = G(n,s) + K\tilde{\Gamma}(-n) F_{\sigma}(n,s) \hat{w}_{n}(s)
  + F(n,s) \hat{h}(-n,s).
\end{equation}
Substituting \eqref{eq:wns} into the above equation, we have
\begin{equation}
\label{eq:general_hat_z}
  \begin{split}
    \hat{z}_{n}(s)
    & = G(n,s) + K\tilde{\Gamma}(-n) \dfrac{F_{\sigma}(n,s) G_{\rho}(n,s)}{D_{K}(n,s)}  \\
    & + \chi(n,s) \hat{h}(-n,s),
  \end{split}
\end{equation}
where the susceptibility $\chi(n,s)$ is
\begin{equation}
  \chi(n,s) = 
  F(n,s) + K\tilde{\Gamma}(-n) \dfrac{F_{\sigma}(n,s) F_{\rho}(n,s)}{D_{K}(n,s)}.
\end{equation}

The weighted-coupling model \eqref{eq:weighted-coupling-model}
in the main text is obtained
by setting $\Gamma(\theta)=-2i\sin(\theta+\alpha)$,
%$\Gamma(\theta)=-\sin(\theta+\alpha)/(2i)$,
which gives $\tilde{\Gamma}(-1)=e^{-i\alpha}$.
Focusing on $n=1$, which corresponds to the Fourier $-1$ mode of $\tilde{f}_{1}$,
and assuming $f_{1}(\theta,\omega,\sigma,\rho,-0)=0$,
we reproduce the linear response formula for the weighted-coupling model.

\section{Analytic continuation}
\label{sec:analytic-continuation}

The functions $F_{X}(s)$ \eqref{eq:FXs}, $F_{X}(n,s)$ \eqref{eq:FXns},
and $G_{X}(n,s)$ \eqref{eq:GXns} are firstly defined in ${\rm Re}\,s>0$,
which is the domain of the Laplace transform \eqref{eq:Laplace-transform}.
We continue these functions into the whole complex $s$ plane,
which is necessary to obtain $F_{X}(i\oex)$ included
in the susceptibility $\chi(i\oex)$, for instance.
We descrive the continuation for $F_{X}(s)$,
but the idea is directly applicable to $F_{X}(n,s)$ and $G_{X}(n,s)$.

In the definition of $F_{X}(s)$,
% $F_{X}(s)$ \eqref{eq:FXs}, % and $G_{X}(s)$ \eqref{eq:GXs}, and of 
% $F_{X}(n,s)$ \eqref{eq:FXns}, and $G_{X}(n,s)$ \eqref{eq:GXns},
the integral with respect to $\omega$ is defined along the contour $L$.
The integral contour $L$ is the real axis for ${\rm Re}\,s>0$ 
and the pole $\omega=-is$ of the integrand is not on $L$.
% Therefore, the integrals can be performed.
% In the limit ${\rm Re}\,s\to+0$, the pole arrives on the real axis,
% and we smoothly modify the integral contour $L$ to avoid the pole.
% The integral contour $L$ is the real axis for ${\rm Re}\,s>0$,
% and the pole $\omega=-is$ is not on $L$.
In the limit ${\rm Re}\,s\to+0$, the pole arrives on the real axis
from the lower side of the complex $s$ plane.
To avoid this pole, we smoothly modify the integral contour $L$
to the upper side, and continue this modification for ${\rm Re}\,s<0$
so that we obtain the continued function $F_{X}(s)$. % and $G_{X}(s)$. 
This continuation gives the explicit form of the integral
over $\omega$ for a regular function $Z(\omega)$ as
\begin{equation}
  \label{eq:paht_integral}
  \begin{split}
    & \int_{L} \dfrac{Z(\omega)}{s-i\omega} d\omega \\
    & = \left\{
    \begin{array}{ll}
      \displaystyle{
        \int_{-\infty}^{\infty} \dfrac{Z(\omega)}{s-i\omega} d\omega },
      & ({\rm Re}\,s>0) \\
      {\rm PV} \displaystyle{
        \int_{-\infty}^{\infty} \dfrac{Z(\omega)}{s-i\omega} d\omega}
      + \pi Z(-is),
      & ({\rm Re}\,s=0) \\
      \displaystyle{
        \int_{-\infty}^{\infty} \dfrac{Z(\omega)}{s-i\omega} d\omega}
      + 2\pi Z(-is),
      & ({\rm Re}\,s<0) \\
    \end{array}
  \right.
  \end{split}
\end{equation}
where % ${\rm PV}$ represents the Cauchy principal value and
the second terms for ${\rm Re}\,s\leq 0$ is caused by
the residue at the pole $\omega=-is$.
%See Appendix ??? for details.
%\red{\blue{\sout{Similar art for the integral path can be found
%in \cite{strogatz-mirollo-matthews-92}.}}}
%\comment{
%Fig.1 of the cited paper represents the integral on the $s$ plane 
%to obtain $R(t)$ from $\hat{R}(s)$.
%So this comment is not suitable.
%}

\section{Ott-Antonsen reduction}
\label{sec:reduction}

We employ the Ott-Antonsen ansatz \cite{ott-antonsen-08,ott-antonsen-09},
which reduces the original system to a low-dimensional system.
The reduction is useful to examine the theory numerically
since the reduced system corresponds to the large population limit.

The Ott-Antonsen ansatz introduce the form of $f$ as
\begin{equation}
  \label{eq:oa_ansatz}
  \begin{split}
    f(\theta,\omega,\sigma,\rho,t)
    = \dfrac{g(\omega,\sigma,\rho)}{2\pi} 
    & \left\{ 1 + \sum_{n=1}^{\infty} \left[
          a^{n}(\omega,\sigma,\rho,t) e^{in\theta}
      \right. \right. \\
    & + \left. \bar{a}^{n}(\omega,\sigma,\rho,t) e^{-in\theta}
      \right] \biggr\},
  \end{split}
\end{equation}
% According to the Ott-Antonsen ansatz we assume that
% \begin{align}
% f\left(\theta,\omega,\sigma,\rho,t\right) =& \frac{g(\omega)p(\sigma|\omega)q(\rho|\omega,\sigma)}{2\pi}\notag\\
% &\biggl[1+\sum_{n=1}^{\infty}\bigl(a^n\left(\omega,\sigma,\rho,t\right)e^{in\theta}\notag\\
% &+\bar{a}^n\left(\omega,\sigma,\rho,t\right)e^{-in\theta}\bigr)\biggr],\label{eq:oa_ansatz}
% \end{align}
where the complex-valued function
%$a\left(\omega,\sigma,\rho,t\right)$ 
$a(\omega,\sigma,\rho,t)$ satisfies the condition $\left\lvert a^n\left(\omega,\sigma,\rho,t\right)\right\rvert<1$ and is regular on the $\omega$-plane.
%As seen Appendix \ref{sec:reduction_derivation},
By using the model equation \eqref{eq:weighted-coupling-model} %and \eqref{eq:Kij}
and the ansatz \eqref{eq:oa_ansatz} we obtain the equation for $a\left(\omega,\sigma,\rho,t\right)$ as
\begin{align}
  \label{eq:oa_a_z}
  \frac{\partial a}{\partial t} =& -i\omega a+\frac{K\sigma}{2}\left(\bar{w}e^{i\alpha}-a^2we^{-i\alpha}\right)\notag\\
  &-\frac{h
  }{2}\left(e^{-i\omega_{\text{ex}}t}-a^2e^{i\omega_{\text{ex}}t}\right),
\end{align}
where the order parameter $w$ depends on $\bar{a}$.

Let us derive reduced equations for the Kuramoto (K) model,
the Sakaguchi-Kuramoto (SK) model,
and the frequency-weighted-coupling (FWC) model.
The concrete forms of $g(\omega,\sigma,\rho)$ are given as
\begin{equation}
  g(\omega,\sigma,\rho) = 
  \left\{
    \begin{array}{ll}
      g(\omega) \delta(\sigma-1) \delta(\rho-1) & \text{(K,SK models)}, \\
      g(\omega) \delta(\sigma-\omega) \delta(\rho-1) & \text{(FWC model)}. \\
    \end{array}
  \right.
\end{equation}
The order parameter $w$ is expressed by
\begin{equation}
  w = \left\{
    \begin{array}{ll}
      \int_{-\infty}^{\infty}d\omega\,g(\omega) \bar{a}(\omega,1,1,t) 
      & \text{(K,SK models)}, \\
      \int_{-\infty}^{\infty}d\omega\,g(\omega) \bar{a}(\omega,\omega,1,t)
      & \text{(FWC model)}, \\
    \end{array}
  \right.
\end{equation}
which is identical with the order parameter $z$
due to the condition $\rho\equiv 1$.
The integration over $\omega$ is performed
by adding the large upper half circle,
which has no contribution to the integral,
and picking up the two poles of $g(\omega)$, \eqref{eq:family},
at $\omega=\Omega+i\gamma_{1}$ and $\omega=-\Omega+i\gamma_{2}$.
The residues give
\begin{equation}
  w(t) = z(t) = k_{1} A(t) + k_{2} B(t),
\end{equation}
where $A$ and $B$ are defined by
\begin{equation}
  A(t) = \left\{
    \begin{array}{ll}
      \bar{a}(\Omega+i\gamma_{1},1,1,t) & (\text{K,SK models}) \\
      \bar{a}(\Omega+i\gamma_{1},\Omega+i\gamma_{1},1,t) & (\text{FWC model}) \\
    \end{array}
  \right.
\end{equation}
\begin{equation}
  B(t) = \left\{
    \begin{array}{ll}
      \bar{a}(-\Omega+i\gamma_{2},1,1,t) & (\text{K,SK models}) \\
      \bar{a}(-\Omega+i\gamma_{2},-\Omega+i\gamma_{2},1,t) & (\text{FWC model}) \\
    \end{array}
  \right.
\end{equation}
and the time-independent coefficients are given by
\begin{equation}
  \begin{split}
    k_{1} & = \dfrac{\gamma_{2}}{\gamma_{1}+\gamma_{2}}
    \dfrac{2\Omega-i\left(\gamma_1+\gamma_2\right)}
    {2\Omega+i\left(\gamma_1-\gamma_2\right)},\\
    k_{2} & = \dfrac{\gamma_{1}}{\gamma_{1}+\gamma_{2}}
    \frac{2\Omega+i\left(\gamma_1+\gamma_2\right)}
    {2\Omega+i\left(\gamma_1-\gamma_2\right)}.
  \end{split}
\end{equation}

Finally, in \eqref{eq:oa_a_z},
setting $\omega$ as $\omega=\Omega+i\gamma_{1}$ or $\omega=-\Omega+i\gamma_{2}$,
and $\sigma=1$ (K,SK models) or $\sigma=\omega$ (FWC model),
we have the reduced equations
\begin{align}
\frac{dA}{d t} =& i\left(\Omega+i\gamma_1\right)A\notag\\
&-\frac{K}{2}\left[A^2\left(\bar{k}_1\bar{A}+\bar{k}_2\bar{B}\right)e^{i\alpha}-\left(k_1A+k_2B\right)e^{-i\alpha}\right]\notag\\
&-\frac{h}{2}\left(A^2e^{-i\omega_{\text{ex}}t}-e^{i\omega_{\text{ex}}t}\right),\label{eq:reduced_eq1}\\
\frac{dB}{d t} =& i\left(-\Omega+i\gamma_2\right)B\notag\\
&-\frac{K}{2}\left[B^2\left(\bar{k}_1\bar{A}+\bar{k}_2\bar{B}\right)e^{i\alpha}-\left(k_1A+k_2B\right)e^{-i\alpha}\right]\notag\\
&-\frac{h}{2}\left(B^2e^{-i\omega_{\text{ex}}t}-e^{i\omega_{\text{ex}}t}\right),\label{eq:reduced_eq2}
\end{align}
for the K and SK models, and
\begin{align}
\frac{dA}{d t} =& i\left(\Omega+i\gamma_1\right)A\notag\\
&-\frac{K}{2}\left(\Omega+i\gamma_1\right)\notag\\
&\left[A^2\left(\bar{k}_1\bar{A}+\bar{k}_2\bar{B}\right)e^{i\alpha}-\left(k_1A+k_2B\right)e^{-i\alpha}\right]\notag\\
&-\frac{h}{2}\left(A^2e^{-i\omega_{\text{ex}}t}-e^{i\omega_{\text{ex}}t}\right),\label{eq:reduced_eq1}\\
\frac{dB}{d t} =& i\left(-\Omega+i\gamma_2\right)B\notag\\
&-\frac{K}{2}\left(-\Omega+i\gamma_2\right)\notag\\
&\left[B^2\left(\bar{k}_1\bar{A}+\bar{k}_2\bar{B}\right)e^{i\alpha}-\left(k_1A+k_2B\right)e^{-i\alpha}\right]\notag\\
&-\frac{h}{2}\left(B^2e^{-i\omega_{\text{ex}}t}-e^{i\omega_{\text{ex}}t}\right),\label{eq:reduced_eq2}
\end{align}
for the FWC model.

\section{Stability analysis by the Nyquist diagram}
\label{sec:Nyquist}

% In the numerical computations, Sec. \ref{sec:numerics},
% we used unimodal $g(\omega)$,
% and hence we review stability analysis by the Nyquist method
% for unimodal distibutions.

For considering the stability of the nonsynchronized state
$f_{0}(\omega,\sigma,\rho)$,
we turn off the external force, $h=0$, and give a small initial perturbation
$f_{1}(\theta,\omega,\sigma,\rho,0)$.
This setting, from \eqref{eq:general_hat_z}, gives the Laplace transform of the order parameter,
$\hat{z}(s)$, as
\begin{equation}
  \label{eq:hatzs-stability}
  \hat{z}(s)
  = G(s) + Ke^{-i\alpha} \dfrac{F_{\sigma}(s) G_{\rho}(s)}{D_{K,\alpha}(s)},
\end{equation}
where $F_{X}(s)=F_{X}(1,s)$, $G_{X}(s)=G_{X}(1,s)$ and $G(s)=G_{1}(s)$.
See \eqref{eq:FXns} and \eqref{eq:GXns} for the definitions of $F_{X}(n,s)$ and $G_{X}(n,s)$.
As discussed in Sec. \ref{sec:linear-response},
the temporal evolution of $z(t)$ is dominated by the roots of $D_{K,\alpha}(s)$,
that is, $f_{0}$ is unstable if there is a root in the region ${\rm Re}\,s>0$.

We focus on the boundary, the imaginary axis ${\rm Re}\,s=0$.
The imaginary axis, $s=iy$ with $y$ real, is mapped
by the mapping $D_{K,\alpha}$ as
\begin{equation}
  D_{K,\alpha}(iy) = 1 - \dfrac{K}{2} e^{-i\alpha}
  \int_{L}d\omega\, \dfrac{g(\omega)}{\omega-y} 
  % \left[
  %   \dfrac{\pi}{2} g(y)
  %   + \dfrac{i}{2} {\rm PV} \int_{-\infty}^{\infty} \dfrac{g(\omega)}{\omega-y} d\omega \right]
\end{equation}
for the Kuramoto and Sakaguchi-Kuramoto models, and 
\begin{equation}
  D_{K,\alpha}(iy) = 1 - \dfrac{K}{2}
  \int_{L}d\omega\, \dfrac{\omega g(\omega)}{\omega-y}
 % \left[
 %    \dfrac{\pi}{2} y g(y)
 %    + \dfrac{i}{2} {\rm PV} \int_{-\infty}^{\infty} \dfrac{\omega g(\omega)}{\omega-y} d\omega \right]
\end{equation}
for the frequency-weighted coupling model.
We remark that the integral can be computed
by referring the continuation discussed in Appendix \ref{sec:analytic-continuation}
and using the residue theorem for the considered $g(\omega)$ \eqref{eq:family}.
The function $D_{K,\alpha}(iy)$ goes to $1$ in the limit $|y|\to\infty$,
and hence, $D_{K,\alpha}$ maps the imaginary axis to a closed circle.
The unstable region, ${\rm Re}\,s>0$, is the right-hand-side
of the imaginary axis oriented from $-i\infty$ to $+i\infty$,
then the right-hand-side of the oriented circle corresponds
to the unstable region.
Therefore, if the right-hand-side includes the origin
of the complex $D_{K,\alpha}(s)$ plane,
there exists a root of $D_{K,\alpha}(s)$ in the unstable region
of the complex $s$ plane.

% To decide the right-hand-side of the circle,
% we study the sign of the imaginary part of $D_{K,\alpha}(iy)$
% in the limits $y\to\pm\infty$.
% In the Kuramoto and the Sakaguchi-Kuramoto models, we have
% \begin{equation}
%   {\rm Im}D_{K,\alpha}(iy)
%   = - \dfrac{K}{2} \left(
%     \cos\alpha {\rm PV} \int_{-\infty}^{\infty} \dfrac{g(\omega)}{\omega-y} d\omega
%     - \sin\alpha \pi g(y) \right).
% \end{equation}
% The first and second terms go to zero in the order of $O(|y|^{-1})$
% and $O(|y|^{-4})$, and the sign of ${\rm Im}D_{K,\alpha}(iy)$ is dominated
% by the factor $(K/y)\cos\alpha$ and 
% \begin{equation}
%   {\rm Im}D_{K,\alpha}(iy) \to \left\{
%     \begin{array}{ll}
%       -0 & (y\to -\infty) \\
%       +0 & (y\to +\infty) \\
%     \end{array}
%   \right.
% \end{equation}
% for $K>0$ and $|\alpha|<\pi/2$. The right-hand-side of the circle
% is, therefore, the inside of the circle.

% In the frequency-weighted coupling model, a similar analysis gives
% \begin{equation}
%   {\rm Im}D_{K,\alpha}(iy) \to \left\{
%     \begin{array}{ll}
%       -({\rm sgn}\ave{\omega}) 0 & (y\to -\infty) \\
%       +({\rm sgn}\ave{\omega}) 0 & (y\to +\infty) \\
%     \end{array}
%   \right.
% \end{equation}

% The considered natural frequency distribution $g(\omega)$

% \textcolor{magenta}{
% it is not obvious the sign of imaginary part of $D$
% except for the Kuramoto model.
% Might it be better to focus on the Kuramoto model?
% }

\begin{figure}
  \centering
  \includegraphics[scale=0.65]{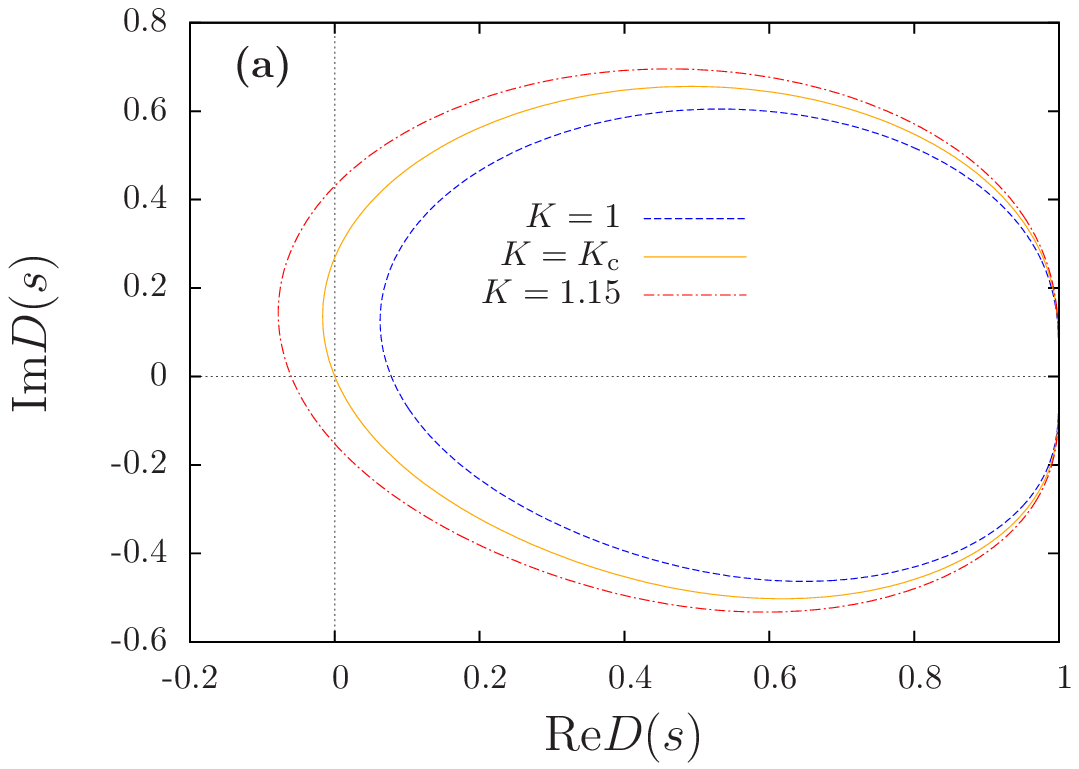}
  \includegraphics[scale=0.65]{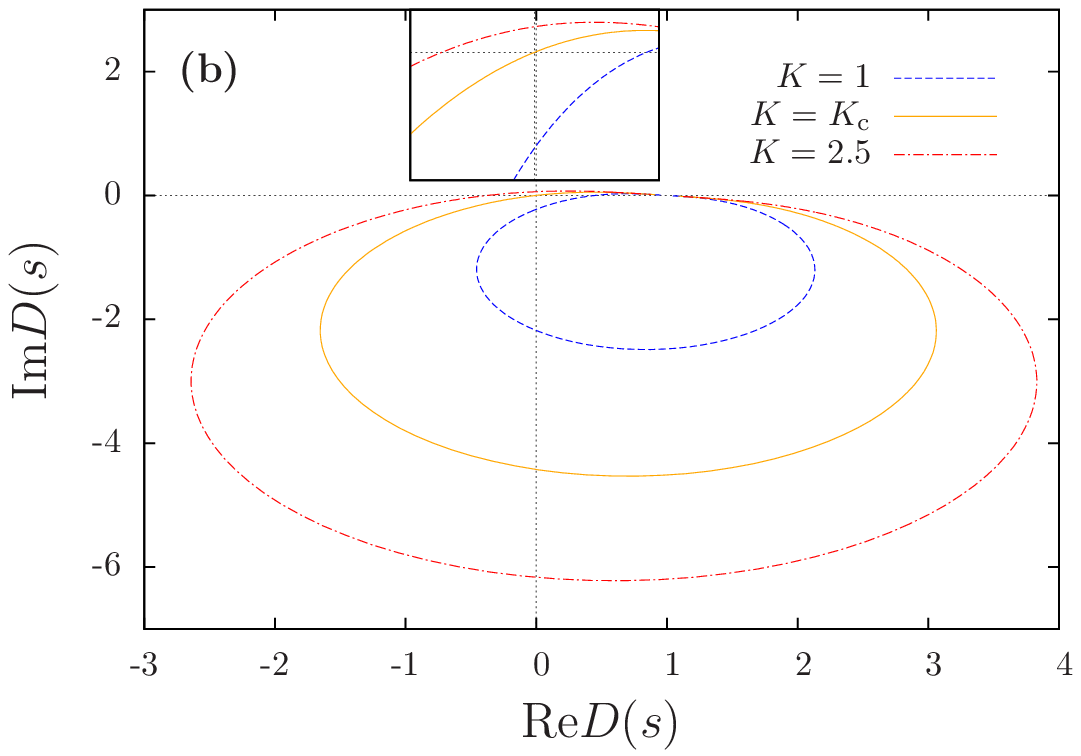}
  \includegraphics[scale=0.65]{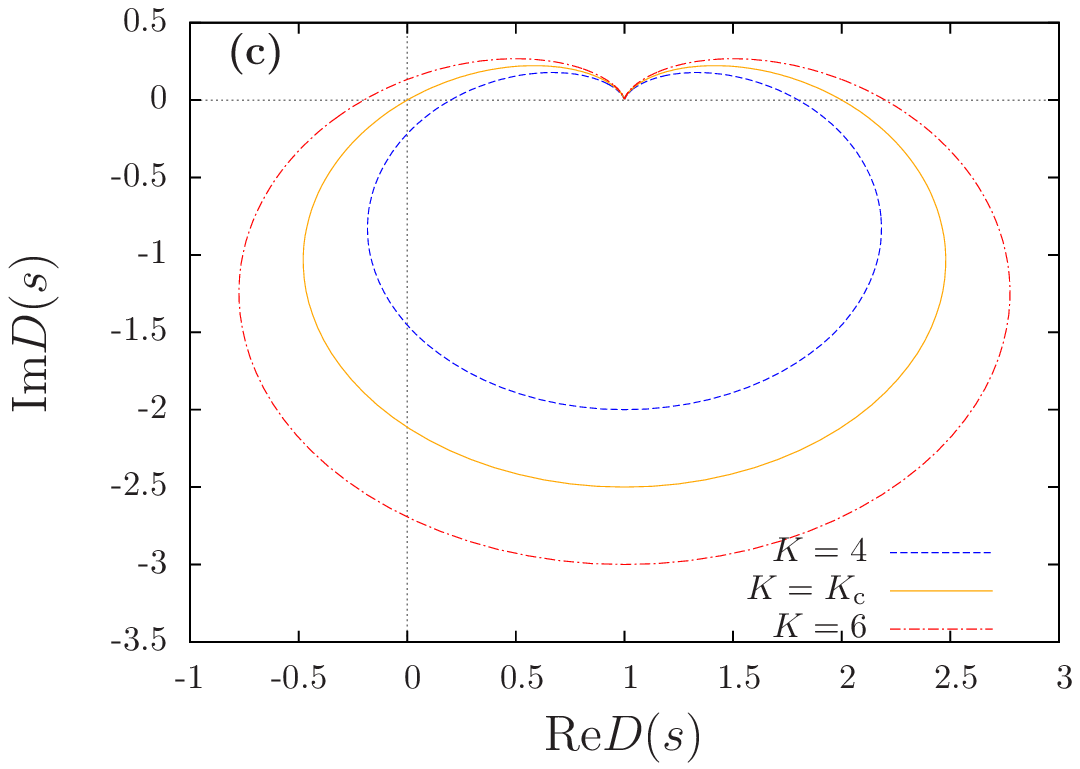}
  \caption{Nyquist diagrams for
    (a) the Kuramoto model with $(\gamma_{1},\Omega)=(0.6,0.6)$ and
    $K_{\rm c}\simeq 1.084618$,
    (b) the Sakaguchi-Kuramoto model with $(\gamma_{1},\Omega)=(0.25,0)$,
    $\alpha=-1.436475$ and $K_{\rm c}\simeq 1.821283$, and
    (c) the frequency-weighted-coupling model with $(\gamma_{1},\Omega)=(0.25,0)$ and
    $K_{\rm c}=5$.
    In each panel three curves are shown for $K<K_{\rm c}$ (blue broken),
    $K=K_{\rm c}$ (orange solid), and $K>K_{\rm c}$ (red dot-dashed).
    The inset of (b) is a magnification around the origin.}
  \label{fig:Nyquist}
\end{figure}

Nyquist diagrams for the three models are described in Fig. \ref{fig:Nyquist},
where the $y$-dependence of $D_{K,\alpha}(iy)$ says
that the right-hand-sides of the circles are their insides.
Note that $K$ modifies the distance from the point $D_{K,\alpha}(iy)=1$
as $D_{K,\alpha}(iy)-1$ is proportional to $K$.
Therefore, in each case, the Nyquist diagram passes
the origin only at the critical $K=K_{\rm c}$
and the nonsynchronized states are confirmed to be unstable for $K>K_{\rm c}$.

\section{Divergence condition in the Kuramoto model with symmetric unimodal distributions}
\label{sec:symmetry-case}

Here, we show that in the Kuramoto model with symmetric unimodal natural frequency distributions the divergence condition $\mathrm{Im}\,F\left(\omega_{\text{ex}}\right)=0$ holds if and only if $\omega_{\text{ex}}=\Omega$, where $\Omega$ is the mode of $g(\omega)$.
We consider $g_{0}(\omega-\Omega)=g(\omega)$,
so that $g_{0}(\omega)$ is even unimodal distribution.
Then, the shifted distribution $g(\omega+\Omega)$ is even.
%We show that the special frequency $\oex$ must be $\Omega$
%to satisfy the condition \eqref{eq:divergence-Kuramoto}.
%\eqref{eq:rotating-ImF}.

We consider the shift of $g(\omega)$ as
$g(\omega+\oex)=g_{0}(\omega+a)$ where $a=\oex-\Omega$
and discuss the function
\begin{equation}
  G = {\rm PV}\int_{-\infty}^{\infty}d\omega \dfrac{g(\omega+\oex)}{\omega}
  = {\rm PV}\int_{-\infty}^{\infty}d\omega \dfrac{g_{0}(\omega+a)}{\omega}.
\end{equation}
From the definition of the even function $a=0$ implies $G=0$.
We show that $G=0$ implies $a=0$
through the contraposition: $a\neq 0$ implies $G\neq 0$. 

Changing the variable $\omega$ to $-\omega$,
the integral is also written as
\begin{equation}
  G = - {\rm PV} \int_{-\infty}^{\infty}d\omega \dfrac{g_{0}(\omega-a)}{\omega}.
\end{equation}
We may therefore assume $a>0$ to prove $G\neq 0$
without loss of generality.
Adding the two expressions, we have
\begin{equation}
  \begin{split}
    2G
    & = {\rm PV} \int_{-\infty}^{\infty}d\omega \dfrac{g_{0}(\omega+a)-g_{0}(\omega-a)}{\omega} \\
    & = \lim_{\epsilon\to +0} ( G^{+} + G^{-})
  \end{split}
\end{equation}
where
\begin{equation}
  \begin{split}
    & G^{+} = \int_{\epsilon}^{\infty}d\omega
    \dfrac{g_{0}(\omega+a)-g_{0}(\omega-a)}{\omega}, \\
    & G^{-} = \int_{-\infty}^{-\epsilon}d\omega
    \dfrac{g_{0}(\omega+a)-g_{0}(\omega-a)}{\omega}. \\
  \end{split}
\end{equation}
From the identity
\begin{equation}
  (\omega+a)^{2}-(\omega-a)^{2} = 4\omega a,
\end{equation}
we can find
\begin{equation}
  \left\{
    \begin{array}{ll}
      |\omega+a| > |\omega-a| & (\text{for } \omega>0), \\
      |\omega+a| < |\omega-a| & (\text{for } \omega<0). \\
    \end{array}
  \right.
\end{equation}
Further, from the unimodality of $g_{0}$, we have
\begin{equation}
  |x| > |y| \quad\Longrightarrow\quad g_{0}(x) < g_{0}(y).
\end{equation}
Putting all together, we have $G^{+}<0$ and $G^{-}<0$,
and hence $G<0$.

\end{document}